\documentclass[preprint,12pt]{elsarticle}

\journal{Computers \& Fluids}
\usepackage{setspace}
\usepackage{graphicx}
\usepackage{caption}
\usepackage{subcaption}
\usepackage{float}
\usepackage{amsfonts} 
\usepackage{changes}
\usepackage{multirow}
\usepackage{tabularx}

\usepackage[nottoc]{tocbibind}
\usepackage{hyperref}
\usepackage[usestackEOL]{stackengine}

\usepackage{amsmath}
\usepackage[linesnumbered,lined,boxed,commentsnumbered]{algorithm2e}
\usepackage{algpseudocode}
\usepackage{babel}

\begin{document}
\begin{frontmatter}

\title{Identification of vortex in unstructured mesh with graph neural networks}

\author[inst1,inst2]{Lianfa Wang}

\affiliation[inst1]{organization={Cemef Mines ParisTech | PSL University},
            addressline={CS 10207 rue Claude Daunesse}, 
            city={Sophia Antipolis},
            postcode={06904}, 
            country={France}}
\affiliation[inst2]{organization={Dpt. Fluid Mechanics, Energy, Environment (MFEE), Electricité de France},
            addressline={6 quai Watier}, 
            city={Chatou},
            postcode={78401}, 
            country={France}}
            
\author[inst2]{Yvan Fournier}
\author[inst2]{Jean-François Wald}
\author[inst1]{Youssef Mesri}

\begin{abstract}
Deep learning has been employed to identify flow characteristics from Computational Fluid Dynamics (CFD) databases to assist the researcher to better understand the flow field, to optimize the geometry design and to select the correct CFD configuration for corresponding flow characteristics. Convolutional Neural Network (CNN) is one of the most popular algorithms used to extract and identify flow features. However its use, without any additional flow field interpolation, is limited to the simple domain geometry and regular meshes which limits its application to real industrial cases where complex geometry and irregular meshes are usually used. Aiming at the aforementioned problems, we present a Graph Neural Network (GNN) based model with U-Net architecture to identify the vortex in CFD results on unstructured meshes. The graph generation and graph hierarchy construction using algebraic multigrid method from CFD meshes are introduced. A vortex auto-labeling method is proposed to label vortex regions in 2D CFD meshes. We precise our approach by firstly optimizing the input set on CNNs, then benchmarking current GNN kernels against CNN model and evaluating the performances of GNN kernels in terms of classification accuracy, training efficiency and identified vortex morphology. Finally, we demonstrate the adaptability of our approach to unstructured meshes and generality to unseen cases with different turbulence models at different Reynolds numbers.
\end{abstract}

\begin{keyword}
Computational Fluid Dynamics (CFD); Vortex identification; Convolutional Neural Network (CNN); Graph Neural Network (GNN); Unstructured mesh.
\end{keyword}
\end{frontmatter}

\section{Introduction}
Vortex identification is of important interest to understand fluid flow characteristics. Deterministic vortex identification methods based on the physical properties of the flow field such as the $Q$-criterion \cite{Q-criterion}, the $\Delta$-criterion \cite{delta-criterion}, and the $\lambda_2$-criterion \cite{lambda-criterion}, provide efficient solutions but with many false positives and false negatives as highlighted in \cite{cons-criterion}. Moreover, these methods require careful selection of appropriate thresholds to obtain valid results. Other methods based on the topological properties of the flow field such as Instantaneous Vorticity Deviation (IVD) \cite{haller2016defining} are more accurate but computationally expensive. Machine learning methods for vortex identification were introduced recently as an alternative to the existing methods and overcome their shortcomings. It is well known that machine learning methods are good at finding the complex non-linear function between the input and the output if a large amount of data is available. And once trained properly they have better generality than a single criterion to flow cases with different conditions and geometries. However, the current applications of machine learning algorithms on vortex detection are limited to regular mesh and simple flow cases.

Recently, many deep learning algorithms have been employed to identify the flow characteristics in CFD databases. The most successful ones are CNNs \cite{CNN-Lecun-98,ye2020flow, monfort2017deep, deng2019cnn, bai2019streampath, berenjkoub2020vortex, strofer2018data, kim2019robust, wang2020vortex}. CNN can effectively learn the local features by implicitly embedding the learned translational-invariant features in the sequence of the learnable parameters in convolution kernels. Different CNN architectures were designed to identify the vortexes, including Eddy-Net \cite{eddy-net}, R-CNN \cite{strofer2018data}, Vortex-Net \cite{deng2019cnn}, Vortex-U-Net \cite{deng2022vortex}, Vortex-Seg-Net \cite{wang2021rapid} and U-Net CNN \cite{berenjkoub2020vortex}. However, the design of the rectangular-like convolution kernel limits the CNNs to only accept the data stored on Cartesian grid which constrains its generality to the data stored on unstructured meshes widely used in the industrial CFD cases. Therefore, most of the current applications of CNNs on detecting flow phenomena are limited to simple flow cases. When dealing with the unstructured meshes and complex geometries, either the mesh deformation \cite{strofer2018data, wang2020vortex, deng2022vortex} or data interpolation \cite{ye2020flow} is used which inevitably introduces numerical errors. 

Compared with CNNs, GNNs taking the data stored on graphs $\mathcal{G}(\mathcal{V}, \mathcal{E})$ as input can naturally adapt to the unstructured meshes used in the real industrial CFD cases to reflect the complex geometries. A graph $\mathcal{G}(\mathcal{V}, \mathcal{E})$ is formed by a set of nodes $\mathcal{V}$ connected by a set of edges $\mathcal{E}$. The data can be stored on the edges, nodes or both. Many GNN kernel functions are proposed to gather neighbor information to the center node. One popular type is the isotropic kernel functions such as graph convolutional network (GCN)\cite{kipf2016semi}, GraphSage \cite{hamilton2017inductive}. The isotropic aggregation functions treat the features from different neighbors equally and fail to take into account the spatial distribution of the neighbors. Another type of GNN kernels is anisotropic and distinguishes the edges by assigning each edge an importance. There are many anisotropic kernels using different mechanisms, like attention mechanism \cite{velivckovic2017graph} or edge gates \cite{bresson2017residual}. However, the above-mentioned kernels are designed to perform regression or classification on networks where there are no Euclidean coordinates, such as social networks, citation networks and biological networks, and only the connections between nodes are important. On the contrary, the coordinates of the nodes or the relative positions of the neighbors to the central nodes are very important on identifying flow characteristics in CFD results which reside on meshes. Therefore, these algorithms are unsuitable to identify the flow phenomena with a certain spatial structure such as the vortex on the CFD meshes.

Inspired by the effectiveness of CNN learning local features in the images, many researchers tried to generalize CNN paradigms to irregular grids which fall into geometric deep learning domain \cite{bronstein2017geometric, bronstein2021geometric}. Masci et al. \cite{masci2015geodesic, masci2015shapenet} designed geodesic convolutional neural networks to extract invariant shape features on Riemannian manifolds by constructing the convolution kernel on a local geodesic system of polar coordinates. Boscaini et al. \cite{boscaini2016learning} generalized convolutions to non-Euclidean domains by constructing a set of oriented anisotropic heat diffusion kernels whose shape, orientation and scale can be varied by the trainable parameters and achieved state-of-the-art results on learning correspondences between deformable shapes. Monti et al. \cite{monti2017geometric} designed a MoNet algorithm which can learn the edge importance according to the pseudo-coordinates of the two connected nodes using the Gaussian Mixture Model (GMM) as the kernel. GMM learns the mean vector and standard deviation of the edge vector pointing from a central node to neighbor nodes. Therefore, GMM assigns an importance to each edge according to the relative position of the connected neighbor nodes with respect to a center node. Fey et al. \cite{fey2018splinecnn} proposed a spline-based convolutional neural network (SplineCNN), a generalization of traditional CNN convolution kernel by using B-spline basis functions parameterized by trainable weights which learns on unstructured and geometric input. To the best of our knowledge, GMM and SplineCNN are the two GNN kernels the most equivalent counterparts of the convolution kernel in traditional CNNs.

In this paper, we propose a GNN framework which can precisely identify the vortex in CFD results as the traditional CNN does but accepts unstructured meshes. The proposed GNN model uses U-Net architecture which utilizes the graph hierarchy generated from AMG method embedded in code\_saturne, an open-source, finite volume method (FVM) based CFD software developed by Electricité de France (EDF), to accelerate the training and improve the classification performance. We detail how both the graphs and graph hierarchy are generated from CFD meshes and propose a vortex auto-labeling method based on biased random walking algorithm to generate dataset. We show both advantages and disadvantages of two GNN kernels, GMM and SplineCNN, the two most equivalent counterparts of the convolution kernel in traditional CNN.

The rest of this paper is organized as follows. The relevant convolution kernels, including traditional CNN, GMM, SplineCNN and GCN, are introduced in Section \ref{relatedwork}. The model architecture, the graph generation details and the algebraic multigrid coarsening method to generate a graph hierarchy for GNN based models are introduced in Section \ref{ArchGraph}. The dataset generation details including CFD case simulation, ground-truth labeling and input sets for machine learning models are introduced in \ref{dataset&input}. The results of a series of trainings, including training details, evaluation of input sets and GNN kernels, GNNs' adaptability to unstructured meshes and generality analysis are shown in Section \ref{results}. Finally, a conclusion is made at the end.

\section{Related work}\label{relatedwork}
During the last several years, CNNs using the convolution to extract the feature embedded have achieved great success in computer vision. Convolution is a specialized type of linear operation used for feature extraction, where a small array of numbers, called a kernel, is applied across the input. An element-wise product between each element of the kernel and the input is calculated at each location of the tensor and summed to obtain the output value in the corresponding position of the output \cite{yamashita2018convolutional}. A typical CNN model normally stacks multiple convolutional layers which use the following equation to update the hidden features from one layer $f^l$ to the next:

\begin{equation}
   f^{l+1}(i, j) = \sigma\left(\Theta^l \cdot (f^l\star g)^l)(i, j)+b^l\right),
\end{equation} 
where $(f \star g)$ represents the convolution kernel, $\Theta \in \mathbf{R}^{C_l\times C_{l+1}}$ represents the trainable mapping hidden features between two consecutive layers and have the dimension of $C_l \times C_{l+1}$, $C$ the channel number, $b \in \mathbf{R}^{C_{l+1}}$ the bias trainable, $\sigma$ the non-linear activation function. 

While the traditional CNN kernel is only suitable for structured data, like images and data on a cartesian mesh, many researchers have tried to extend its efficiency and accuracy to graph domain including GMM, SplineCNN, GCN and so on. In this section, we introduce the details of these different convolution kernels.

\noindent\textbf{Traditional CNN.} In traditional CNN, the convolution between the kernel of the size $H\times W$ and the input data located at point $(i, j)$ in the receptive field or image can be expressed as follows:
\begin{equation}
   (f\star g)(i, j) = \sum_{h=1}^H\sum_{w=1}^W f(i+h, j+w) \cdot g_{h, w},
\end{equation}
where $g$ is a rectangular trainable kernel with dimension of $H\times W$. As shown in this equation, each neighbor pixel always times the element at the same position in the convolution kernel according to their position relative to the center pixel which enables CNNs to know where the information comes from. And the values in the learned kernels represent the importance of the information from the corresponding directions. Therefore the convolution kernel in traditional CNNs associating the direction with the importance of each gathered features is the key to its success.

\noindent\textbf{GMM.} Monti et al. \cite{monti2017geometric} extended the traditional CNN convolution to graph by using Gaussian Mixture Model. GMM kernel gathers the hidden features from neighbors to center node with the following equation:
\begin{equation}
    (f\star g)(i) =\frac{1}{K}\sum_{k=1}^K \frac{1}{|\mathcal{N}(i)|}\sum_{j \in \mathcal{N}(i)}f(j) \cdot g_k (\mathbf{e}_{ij}),
\end{equation}
where hyperparameter $K$ represents the number of directions learned in the kernel, $|\mathcal{N}(i)|$ the degree of node $i$, the edge $\mathbf{e}_{ij}$ pointing from central node $i$ to neighbor node $j$:
\begin{equation}
    \mathbf{e}_{ij}=\left[x_j, y_j\right]^T - \left[x_i, y_i\right]^T,
\end{equation}
and $g_k (\mathbf{e}_{ij})$ the alignment between the edge $\mathbf{e}_{ij}$ and the $k$-th direction $\mu_k$ in the kernel:
\begin{equation}
    g_k (\mathbf{e}_{ij}) =\exp\left[-\frac{(\mathbf{e}_{ij} -\mathbf{\mu}_k)^T (\mathbf{e}_{ij}- \mathbf{\mu}_k)} {2\sigma_k}\right], k \in \left[1, 2,..., K\right],
\end{equation}
where two trainables $\mu_k$ and $\sigma_k$ represent the $k$-th learned direction in the kernel and its variance, respectively. The better the edge $\mathbf{e}_{ij}$ and the learned direction $\mu_k$ aligned, the higher $g_k (\mathbf{e}_{ij})$.

\noindent\textbf{SplineCNN.} Fey et al. \cite{fey2018splinecnn} generalized the traditional CNN convolution kernel to interpolate the edge importance from fixed positions to desired positions using B-spline basis functions with the following equation:
\begin{equation}
    (f\star g)(i) = \frac{1}{|\mathcal{N}(i)|}\sum_{j\in \mathcal{N} (i)} f(j) \cdot  \sum_{\mathbf{k} \in \mathcal{K}} B_\mathbf{k} (\mathbf{e}_{ij}) \cdot g_\mathbf{k},
\end{equation}
where $\mathcal{K}$ is the Cartesian product of the B-spline bases:

\begin{equation}
     \mathcal{K}=N^{1}_{k_1, p}\times... \times N^{D}_{k_D, p}, \  \mathbf{k}= (K_1, ... , K_D),
\end{equation}
$\mathbf{k}$ represents the number of control points on each dimension of the $D-$ dimensional kernel, and $g_\mathbf{k}$ is the trainables and the control points as well associated with the corresponding product $B_\mathbf{k}(\mathbf{e})$ of the basis functions in $\mathcal{K}$:

\begin{equation}
    B_\mathbf{k}(\mathbf{e}) = \Pi_{d=1}^D N_{k_d, p}^d (e^d).
\end{equation}
The $k$-th B-spline basis function of degree $p$, written as $N_{k, p}(e^d)$, is defined recursively as follows:
\begin{equation}
    N_{k, 0}(e^d) = 
    \left\{
        \begin{array}{cc}
        1, & \mbox{if $u_k \le e < u_{k+1}$}. \\
        0, & \mbox{otherwise}. \\
        \end{array}  
    \right.
\end{equation}

\begin{equation}
    N_{k, p}(e^d) = \frac{e-u_k}{u_{k+p}-u_k}N_{k, p-1}(e^d)+ \frac{u_{k+p+1}-e^d}{u_{k+p+1}-u_k}N_{k+1, p-1}(e^d).
\end{equation}
where $u_k$ is the knot vector and $e^d$ is the edge coordinate on $d$-th dimension.

\begin{figure}[!ht]
\centering
     \begin{subfigure}{\textwidth}
         \centering
         \includegraphics[scale=0.18]{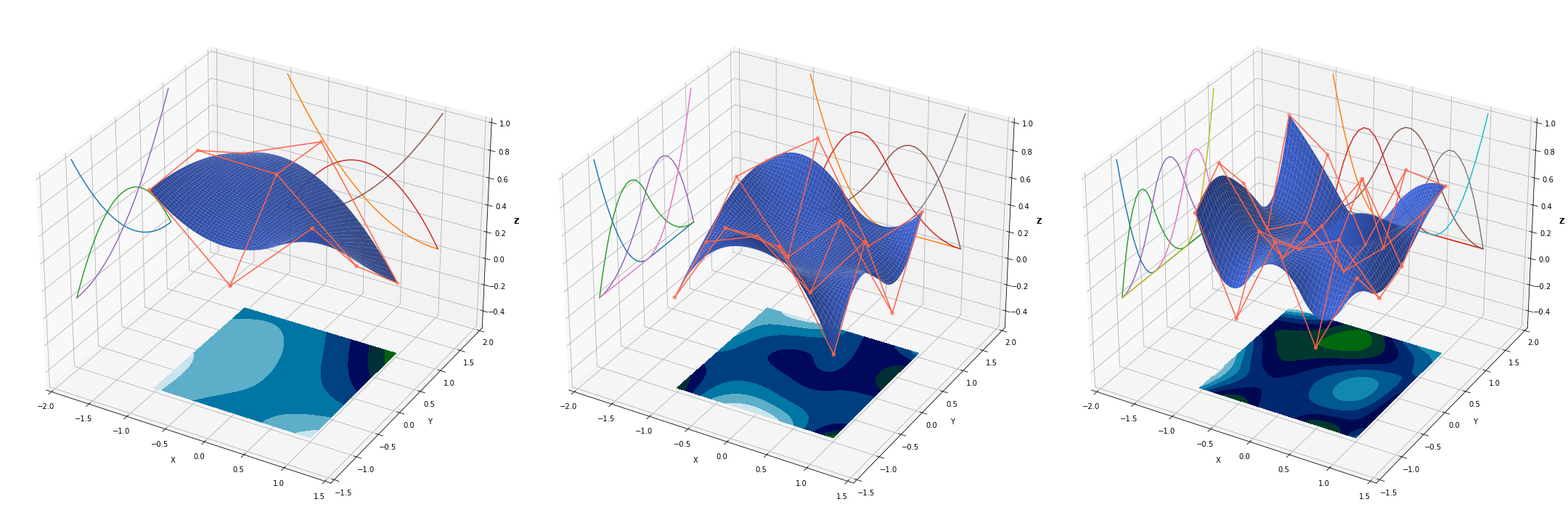}
     \end{subfigure}
\caption{Examples of the convolution kernels for B-spline basis degrees $p=2$ with control points number $K_1=K_2=$ 3, 4 and 5, from left to right. The orange points connected by orange wireframe represent the randomly sampled control points. The curves on the yz-plane and xz-plane represent basis functions for each control point on corresponding direction.}
\label{Fig.B-spline-surface}
\end{figure}

Different from those used in \cite{fey2018splinecnn}, multiple knots are used at the ends of x and y directions to force the interpolated surface to converge to the control points at the ends. Three example convolution kernel surfaces are visualised in Fig. \ref{Fig.B-spline-surface}.

\noindent\textbf{GCN.} A GCN \cite{kipf2016semi} kernel gathers information from all neighbors $\mathcal{N}(i)$ to the center node $i$ with the following equation:
\begin{equation}
    (f\star g)(i) =\frac{1} {\sqrt{|\mathcal{N}(i)||\mathcal{N}(j)|}} \sum_{j \in \mathcal{N}(i)} f(j) \cdot g_j
\end{equation}
where $\mathbf{g}$ is a $\mathcal{N}(i)$ unit vector. GCN kernel is an isotropic kernel and permutation invariant which means the information from all the neighbors is equal no matter in what sequence or from which direction it comes. Therefore, a bad performance of GCN on feature extraction is anticipated. However it shows us where the low bound of the classification performance of a GNN-based algorithm can be, thus GCN kernel is also included in the comparison.

\section{Architecture and graphs}\label{ArchGraph}
\subsection{Architecture}
\begin{figure}[!ht]
\centering
\includegraphics[scale=0.42]{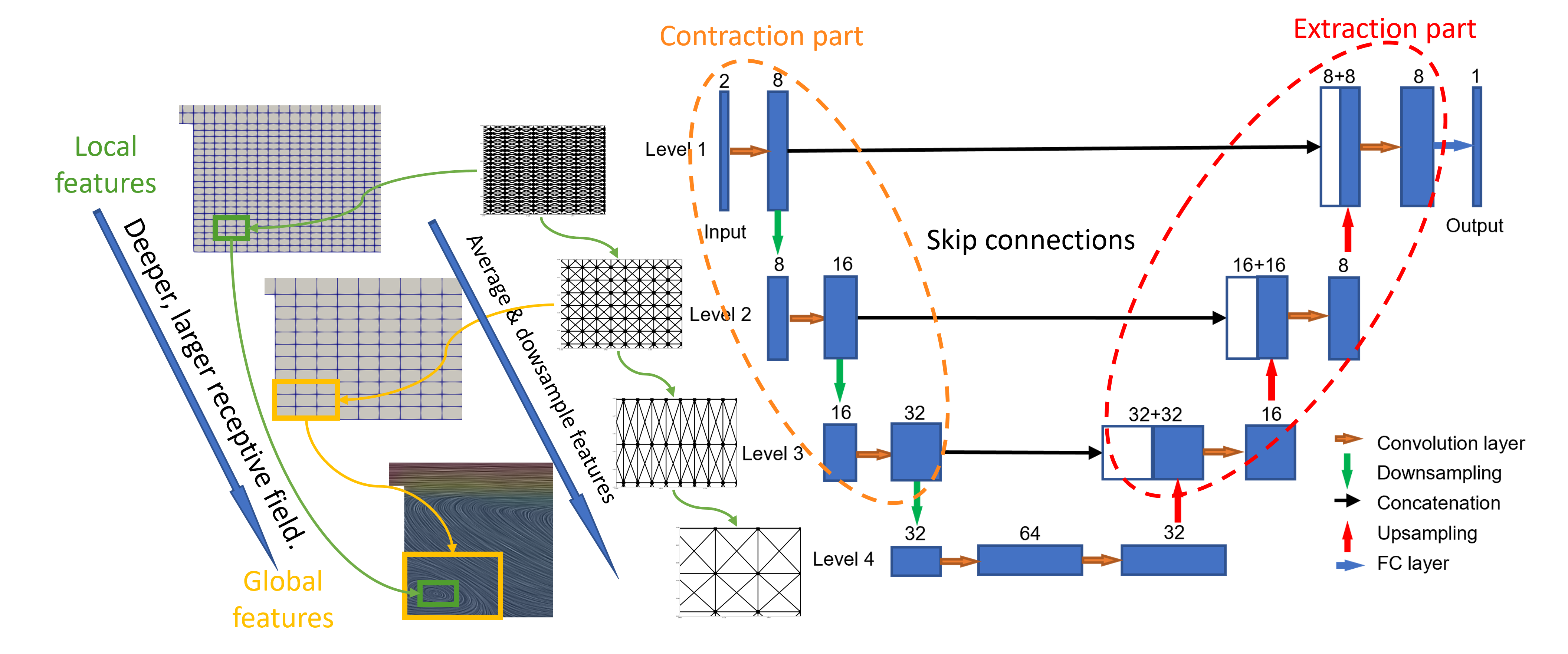}
\caption{U-Net architecture.}
\label{Fig.UNetArch}
\end{figure}

A 4-depth and 8-layer U-Net architecture, as shown in Fig. \ref{Fig.UNetArch}, is adopted to evaluate all convolution kernels. It consists of a leading contract part, a trailing expansive part, the skip-connections among the corresponding contract and expansive parts, and a bottom level. The contract part has three blocks with each of them having one convolutional layer and a following down-sampling (average pooling) layer. In the contract part, the convolutional layer doubles the channel number except that of the first depth level where the output channel is fixed to 8 and the input channel is varied, and the down-sampling layer shrinks the hidden feature size from one level to the next deeper level. Correspondingly, the expansive part also has three blocks with each of them having one up-sampling layer and one following convolutional layer. In expansive part, the up-sampling layers up-sample hidden features from a deeper level to the shallower level, and the convolutional layers decrease the channel number by a factor of 4. Each convolutional layer is followed by a rectified linear unit (ReLU) activation function. The successive down-samplings of the input not only reduce the input size thus leading to computational efficiency, but also enable the U-Net architecture to extract local features and global features in the input since a kernel of the same size at deeper levels can cover larger region with respect to the original flow field. The skip-connections alleviate the vanish gradient problem in the sequentially stacked deep architectures. A fully connected (FC) layer is added to the end to map the output to the classification. 

\subsection{Graph generation}
\begin{figure}[!ht]
    \centering
    \includegraphics[scale=0.25]{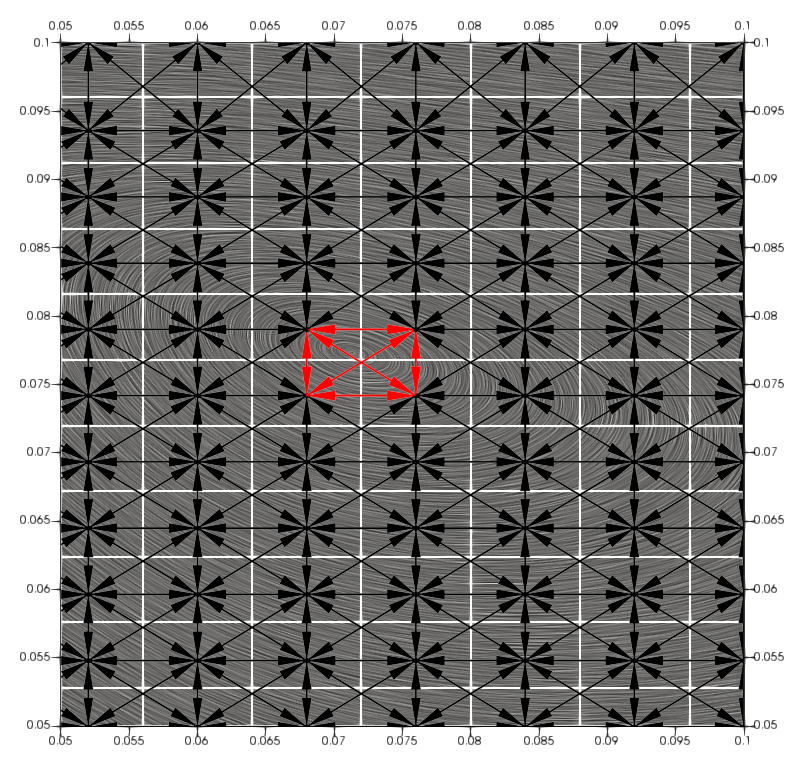}
    \caption{Bi-directed graph superimposed on streamline background. The red arrows connect the vortex core cells. The self-loop of each node is not show here. The white wire frame is the CFD mesh.}
    \label{fig:BidirectedGraph}
\end{figure}

The graphs fed to the GNNs are derived from the CFD meshes where the cells and the vertices shared by adjacent cells in CFD meshes become the nodes and the edges in graphs. The derived graph is a dual mesh of the CFD mesh. The input data are stored on the nodes. To regularize the graph, the normalized direction $\mathbf{e}_{ij}$ pointing from center node $i$ to neighbor node $j$ is stored on the edge connecting the two nodes:

\begin{equation}
    \mathbf{e}_{ij} = \frac{\mathbf{x}_j-\mathbf{x}_i} {||\mathbf{x}_j-\mathbf{x}_i||}
\end{equation}
where, $\mathbf{x_i}$ and $\mathbf{x_j}$ are the center node and neighbor node coordinates, respectively. There exists one forward edge and one backward edge between two connected nodes since every node can be a center node and a self-loop is added to each node with the distance of $\mathbf{0}$. Therefore the bi-directed graphs with self-loops are generated, as shown in Fig. \ref{fig:BidirectedGraph}.

\subsection{Graph coarsening}
\begin{figure}[!ht]
\centering
    \begin{subfigure}{1\textwidth}
         \centering
         \includegraphics[scale=0.18]{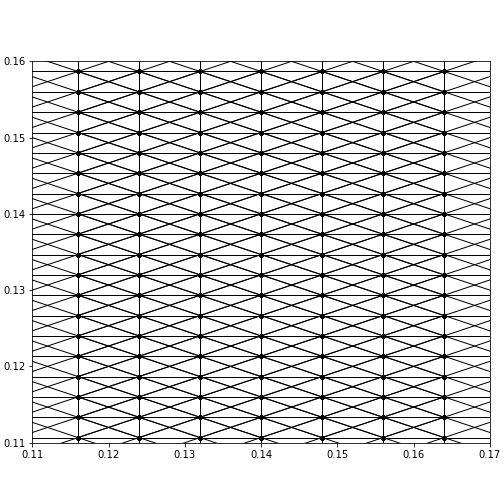}
         \includegraphics[scale=0.18]{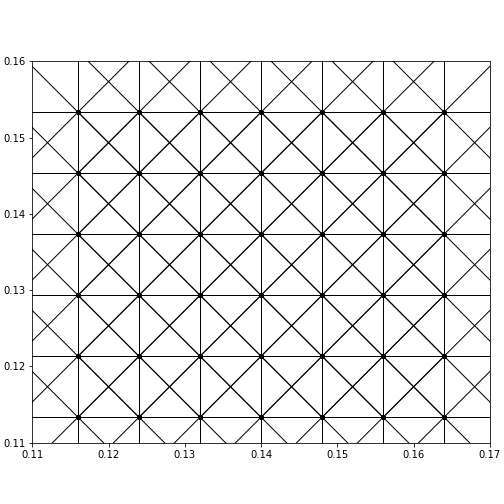}
         \includegraphics[scale=0.18]{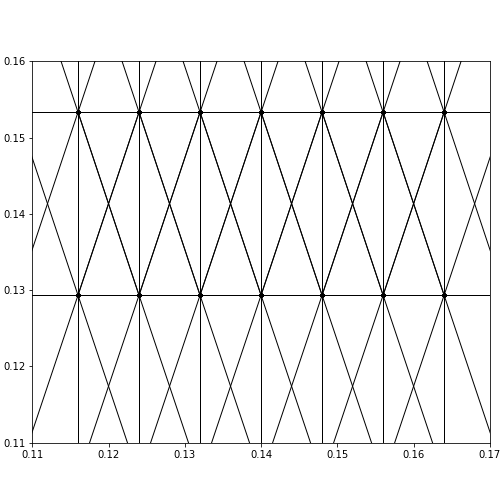}
         \includegraphics[scale=0.18]{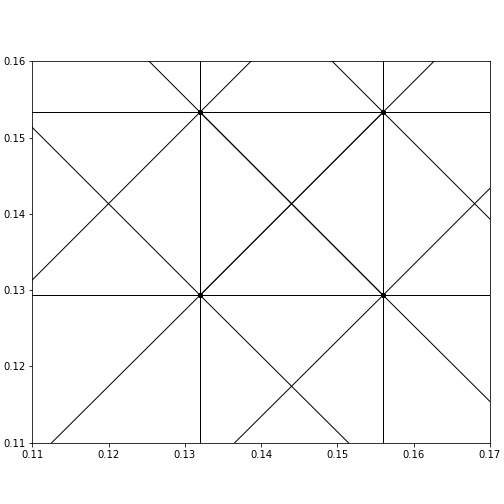}
         \caption{Structured mesh.}\label{Fig.AMG_Graphs_structured}
    \end{subfigure}
    \begin{subfigure}{1\textwidth}
         \centering
         \includegraphics[scale=0.18]{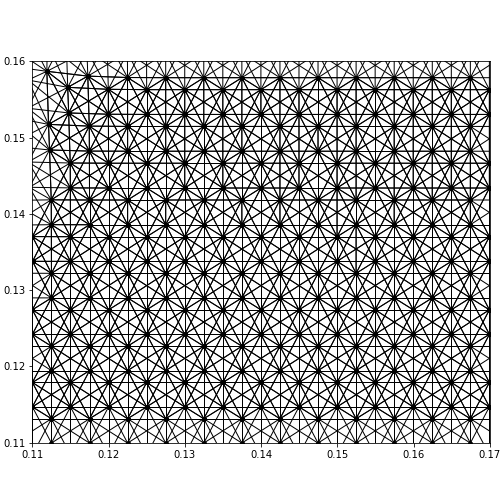}
         \includegraphics[scale=0.18]{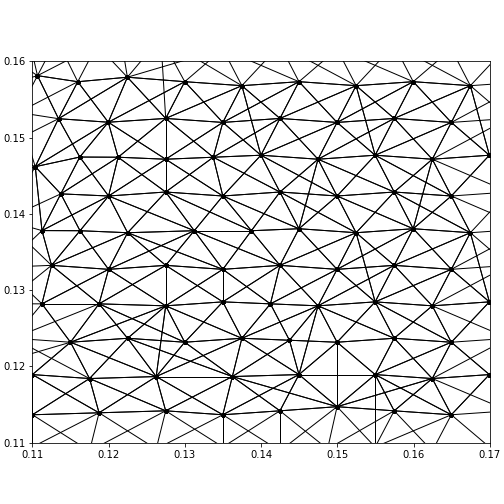}
         \includegraphics[scale=0.18]{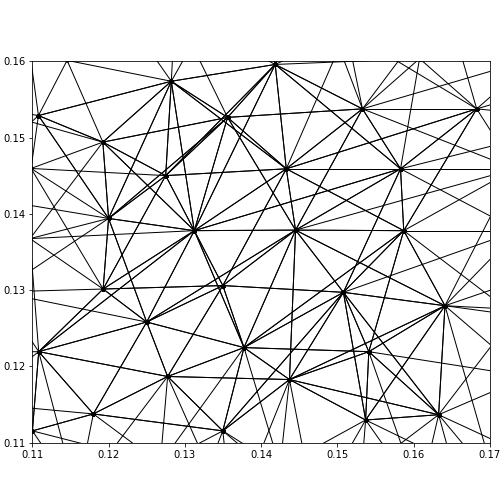}
         \includegraphics[scale=0.18]{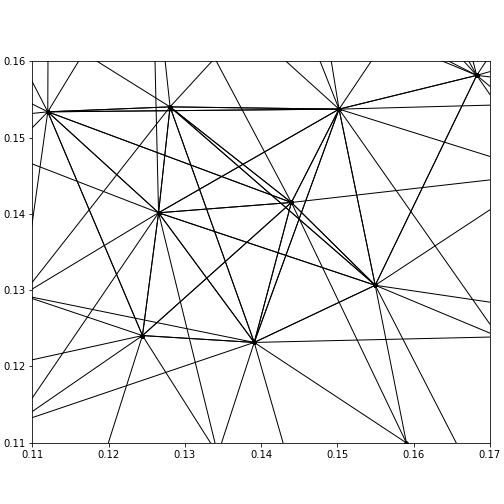}
         \caption{Unstructured mesh.}\label{Fig.AMG_Graphs_unstructured}
    \end{subfigure}
    \caption{Graph hierarchy generated from BFS meshes for four levels in U-Net architecture. The first graph is generated from original mesh, the following three graphs are generated by AMG.}\label{Fig.BFS_graph_herarchy}
\end{figure}

On the contrary to the widely used pooling methods in CNNs thanks to their simplicity, there is no widely-accepted sampling method for GNNs because it requires a hierarchy of graphs with different refinement levels whose generation depends on their application domain. There are several different methods to coarsen graphs including but not limited to: 1) spectral graph sparsification \cite{shuman2015multiscale}, 2) algebraic multigrid (AMG) \cite{notay2010aggregation, edwards2016graph}, 3) geometric graph coarsening and even 4) GNN-based graph coarsening \cite{cai2021graph}. In code\_saturne, an algebraic multigrid method is used to coarsen the meshes to accelerate the convergence of the linear algebraic equations $Au=b$. The equations resulted from fine grid are iterated once and the residuals are restricted to right hand side of the equations at the next level where a coarser grid is constructed by aggregating the strongly coupled adjacent cells at fine levels which is decided by the relative magnitude of the corresponding values in the coefficient matrix. This process can go on and on until a maximum number of coarsening levels is reached or a matrix is small enough to be directed solved by conjugate gradient iteration method. Then the converged solution is successively prolonged from coarser levels to finer levels to correct the solution in the finer level. AMG and U-Net architecture are both designed to separate components of different scales in the signal in order to accelerate the calculation. We construct the graph hierarchy based on the coefficient matrix from the pressure correction step at this stage. In the pressure correction step, the Poisson equation is solved: 
\begin{equation}
    div(\Delta t \underline{\nabla} p') = div(\rho \underline{\tilde{u}})
\end{equation}
where $\Delta t$ is the time step, $p'$ the pressure increment and $\underline{\tilde{u}}$ the velocity field resulting from the prediction step, $\rho$ density. For the internal node $i$, the integrated form of the above equation is:
\begin{equation}
    \sum_{j\in Neigh(i)}[\Delta t(\underline{\nabla}p')_{f_{ij}}]\cdot S_{ij} = \sum_{j\in Neigh(i)}\rho \underline{\tilde{u}}\cdot \underline{S}_{ij}
\end{equation}
where $j$ is $i$'s neighbor node, $f_{ij}$ the interface between nodes $i$ and $j$, and $S$ the surface vector. Since the AMG method used here is aggregation-based which uses only the information present in the system matrix $A$, which here corresponds to $\Delta t S_{ij}$. It means that the mesh coarsening depends only on the mesh topology which can evenly coarsen the original mesh to much deeper levels. The coarsening based on other variables will be tested in the future. The details of the AMG algorithm are out of the scope of this paper and the interested readers are referred to Y. Notay's work\cite{notay2010aggregation}. Two example graph hierarchies generated using AMG method from backward-facing step (BFS) structured and unstructured meshes are shown in Fig. \ref{Fig.BFS_graph_herarchy}. The sizes of images used for CNNs and graphs generated by AMG from different mesh types are summarized in Table \ref{table.GraphsDetails}. Since CNNs only accept a rectangular array, therefore only 240$\times$220 mesh cells behind the step are used as input. By contrary, the graph derived from the entire mesh is used for GNN-based methods, thus leading to the discrepancy between the number of pixels in images and that of nodes in graphs at level 1. 

\begin{table}[ht]
\caption{The details of images and graphs.}
\centering
\begin{tabularx}{0.8\textwidth}{ c | c | c c | c c }
\hline
\multirow{2}{*}{Level} & \multirow{2}{*}{\Centerstack{Imges \\$(Pixels_{W\times H})$ }}    &  \multicolumn{2}{c|}{\Centerstack{AMG graphs \\ (Structured)}} & \multicolumn{2}{c}{\Centerstack{AMG graphs\\ (Unstructured)}} \\
\cline{3-4}\cline{5-6}
          &                   & Nodes & Edges & Nodes & Edges\\
        \hline
        1 & $52800_{240\times220}$ & 74400 & 666124 &  131118  & 1693758\\
        2 & $13200_{120\times110}$ & 24720 & 220002 &   50780  &  441429\\
        3 & $3300_{60\times55}$    & 8231  & 72401  &   17127  &  196325\\
        4 & $810_{30\times27}$     & 2719  & 23727  &    5596  &   67212\\
        \hline
\end{tabularx}
\label{table.GraphsDetails}
\end{table}

\section{Dataset}\label{dataset&input}
\subsection{CFD case simulation}
The dataset was generated with code\_saturne \cite{saturnetheory} which is an open-source software developed primarily by EDF for CFD applications. It solves the Navier-Stokes equations for 2D, 2D-axisymmetric and 3D flows, steady or unsteady, laminar or turbulent, incompressible or weakly dilatable, isothermal or not, with scalar transport. It provides multiple turbulence models, including Reynolds-Averaged Navier-Stokes (RANS) models, Reynolds Stress Models (RSM) and Large Eddy Simulation (LES) models, and many of specific physical modules: coal and heavy-fuel oil combustion, semi-transparent radiative heat transfer, particle-tracking with Lagrangian modeling, Joule effect, electrics arcs, weakly compressible flows, atmospheric flows, rotor/stator interaction for hydraulic machines.

The CFD simulation of turbulent flow over BFS \cite{le1997direct} is simulated using transient code\_saturne solver by solving Navier-Stokes equations:

\begin{eqnarray}
&\displaystyle\frac{\partial \rho}{\partial t}+div(\rho \underline{u})&=0 ,\\
&\displaystyle\frac{\partial \rho\underline{u}}{\partial t}+div\left(\rho\underline{u} \otimes \underline{u}+P\underline{\underline{Id}}+\underline{\underline{R}}\right)&=0,
\end{eqnarray}
and $R_{ij}-\epsilon$ turbulence model \cite{archambeau2004code}:

\begin{eqnarray}
&\displaystyle\frac{\partial R_{ij}}{\partial t}+div\left(\rho\underline{u} R_{ij}-\mu \underline{\nabla }R_{ij}\right)&=P_{ij}+G_{ij}+\Phi _{ij}+d_{ij}-\rho\epsilon_{ij}+S'_{ij},\\
&\displaystyle\frac{\partial \epsilon}{\partial t}+div\left(\rho\underline{u} \epsilon-\mu \underline{\nabla }\epsilon\right)&=d_\epsilon +C_{\epsilon 1}\frac{\epsilon}{k}P-\rho C_{\epsilon 2}\frac{\epsilon^2}{k}+S'_\epsilon,
\end{eqnarray}
where, $P_{ij}, G_{ij}, \Phi_{ij}, d_{ij}, \rho\epsilon_{ij}$ are the generation term, production-destruction term related to gravity effects, pressure-strain term, dissipation term and turbulent diffusion term of $R_{ij}$ respectively, $d_\epsilon$ and $P$ turbulent diffusion term and generation term for $\epsilon$ respectively, $S_{ij}$ and $S_\epsilon$ additional source terms for $R_{ij}$ and $\epsilon$ respectively.

\begin{figure}[!ht]
\centering
\includegraphics[width=0.8\textwidth]{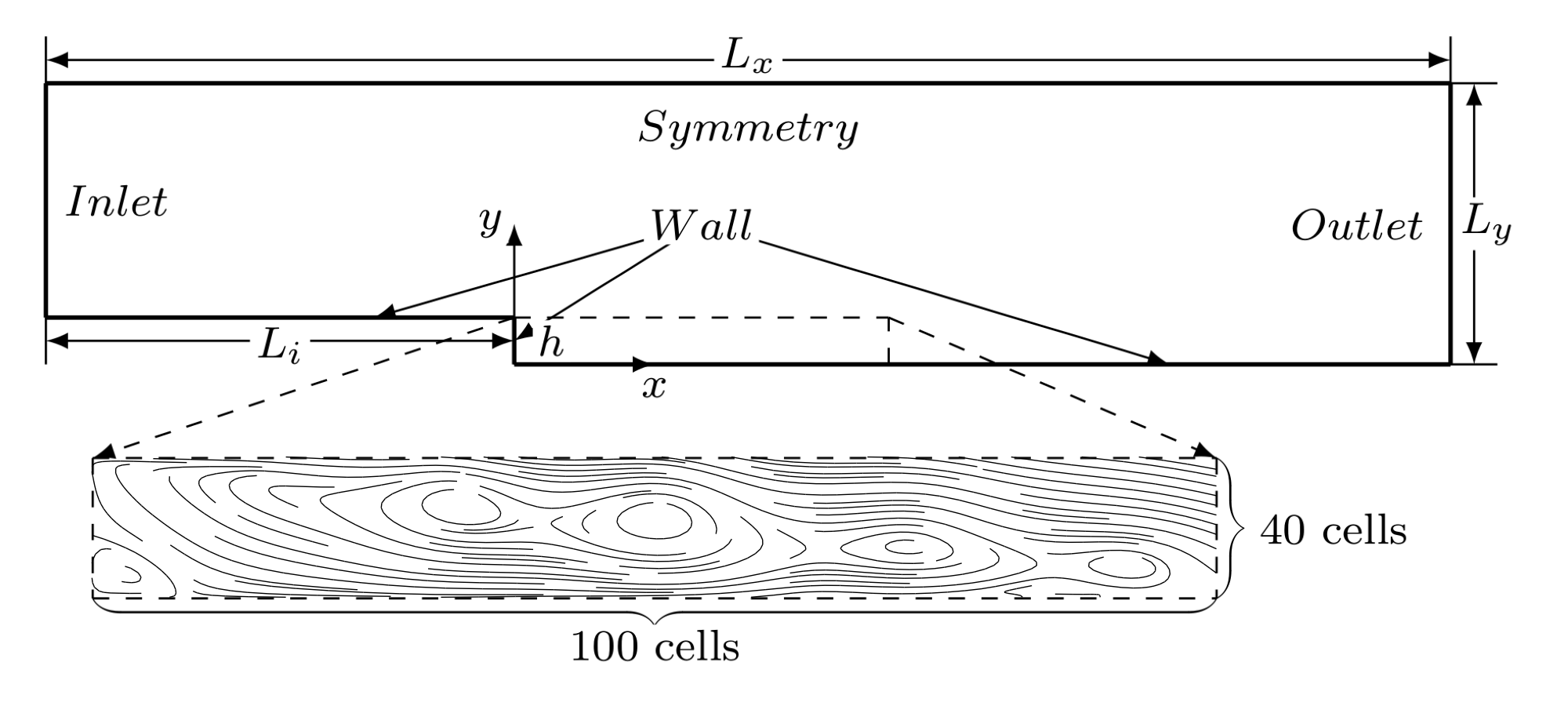}
\caption{Backward-facing step flow configuration.}
\label{Fig.bfs}
\end{figure}

The flow configuration is shown in Fig. \ref{Fig.bfs}. The computational domain consists of an inlet section $L_i=10h$ prior to the sudden expansion. The total length in streamwise direction is $L_x=30h$ and the vertical height is $L_y=6h$. The mean velocity with the turbulence intensity of $5\%$ of the mean velocity magnitude is imposed at the inlet boundary. The Neumann condition is imposed at the outlet boundary. The scalable wall function is applied to the bottom wall. The upper boundary is set to symmetry boundary condition. The Reynolds number $Re_h$ based on step height $h$ and bulk velocity is 5100. The total number of computational cells in x direction after the step is 240 and that in y direction is 220. Only one cell is used for z direction.

At least twenty flow-throughs, flow time from 0s to 10s, were calculated before outputting the result in order to make sure the flow field fully developed. Then the physical fields at 100 time steps with an interval of 0.05s, flow time from 10s to 15s, were output and divided into training, validation and testing cases . The dataset contains results at 100 time steps in total of which the first 80 times steps from 10s to 14s, 10 time steps from 14s to 14.5s, and 10 time steps from 14.5s to 15s were used as training cases, validation cases and test cases, respectively.

\subsection{Ground-truth labeling}
\begin{figure}[!ht]
    \centering
    \includegraphics[scale=0.25]{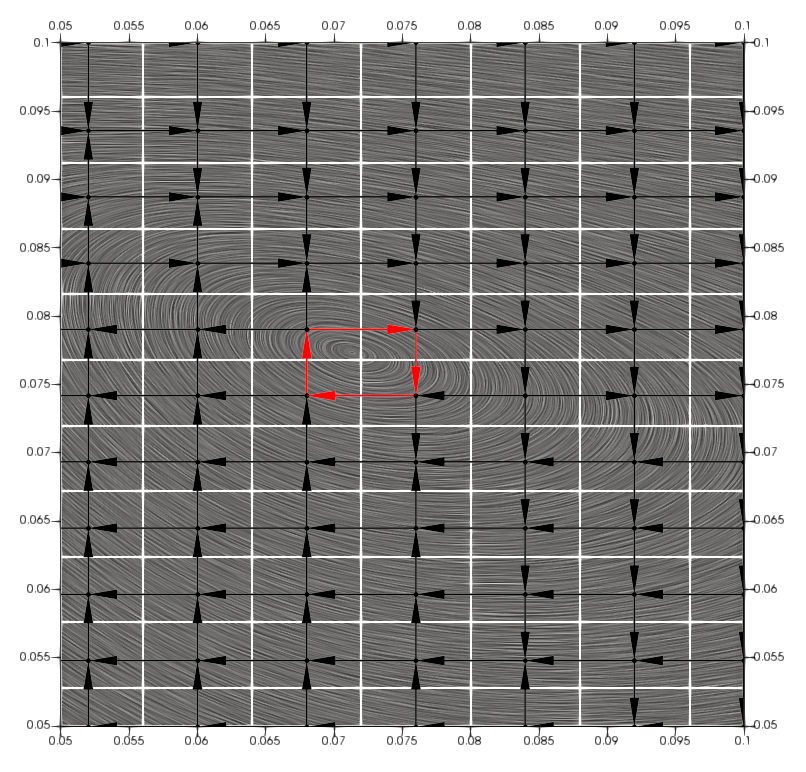}
    \caption{Directed graph superimposed on streamline background. The red arrows connect the vortex core cells. The white wire frame is the CFD mesh.}
    \label{fig:DirectedGraph}
\end{figure}

Since the supervised training of the neural networks is adopted in our study, a binary ground-truth label for each node indicating whether the node is in the vortex region or not should be provided. We propose an auto-labeling method using Depth First Search (DFS) and Biased Random Walking (BRW) on the directed graph derived from the velocity field, as shown in Fig. \ref{fig:DirectedGraph}. The edge in the directed graph is the shared face of adjacent cells in the mesh whose direction is from upwind cell to downwind cell according to the velocity field. 

\begin{figure}[!ht]
    \centering
     \begin{subfigure}{1\textwidth}
         \centering
         \includegraphics[scale=0.23]{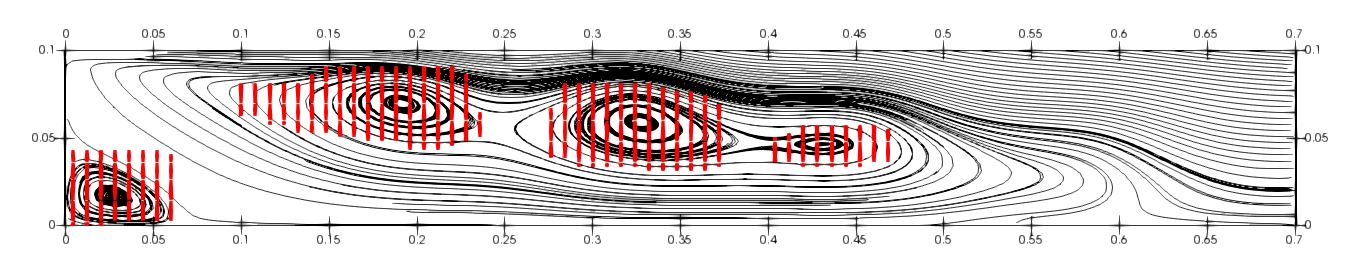}
         \caption{Structured mesh.}
     \end{subfigure}
    \vfill\centering
     \begin{subfigure}{1\textwidth}
         \centering
         \includegraphics[scale=0.23]{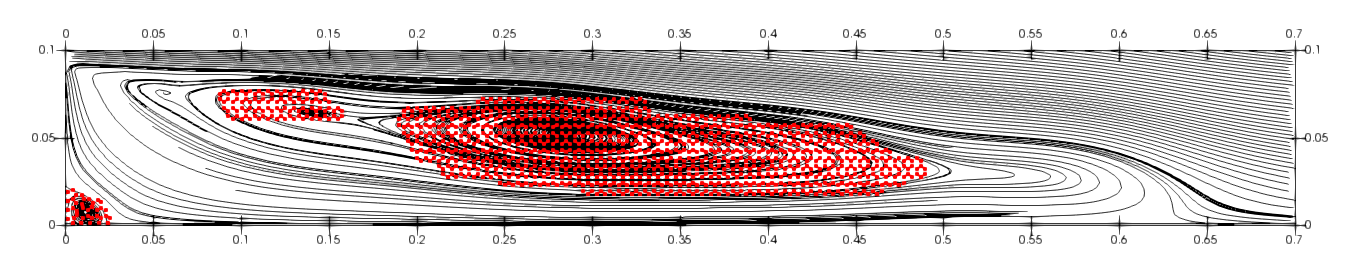}
         \caption{Unstructured mesh.}
     \end{subfigure}
    \caption{Ground-truth label points on streamline background.}
    \label{Fig.ground-truth}
\end{figure}

As shown in Algorithm \ref{alg:autolabeling}, this vortex auto-labeling algorithm labels the vortex in 2D CFD results based on the closure of the streamline in two steps: a) locating the vortex core with DFS algorithm \cite{jungnickel2005graphs}; b) enlarging the vortex region by BRW. For the first step, we assume that a vortex core exists among limited adjacent cells. The DFS algorithm, as shown in Algorithm \ref{alg:DFS}, is used to traverse all the nodes on the graph and recursively search the downstream nodes until it goes back to the beginning node within predefined steps. Once the vortex core is located, the BRW algorithm, as shown in Algorithm \ref{alg:BRW}, enlarges the vortex region starting from the vortex core. The probability of random walking from central node to a downstream node is proportional to the ratio of mass flux entering the corresponding downstream node from central node to the total mass flux exiting the central node. One walking path either ends at the beginning node leading to the success of enlarging the vortex region or ends at the outlet nodes leading to the failure of enlarging vortex region. To make sure the final labeled vortexes are separate from each another, a single enlarged vortex region shouldn't contain multiple vortex cores. The vortex location labeling process is terminated when the searched vortex boundary is not changed from the last random walking. The auto-labeling algorithms takes 292s and 122s on average to label vortexes on the 2D BFS structured and unstructured meshes, respectively. The streamline plot and ground-truth label obtained with this method at one time instant from both structured and unstructured meshes are shown in Fig. \ref{Fig.ground-truth}. All the cells in vortex zone are represented by the points over the streamline background. It should be noted that although the vortex boundary labeled in this method doesn't precisely reflect the real vortex shape leading to a noisy dataset, it has the merit of auto-labeling and can label vortex location for massive snapshots. And we expect that the machine learning algorithm outperforms the ground-truth label on identifying the vortex position and shape after trained on this noisy dataset.

\SetKwComment{Comment}{/* }{ */}
\SetKwInOut{Input}{input}
\SetKwInOut{Output}{output}
\SetKwFunction{DFS}{DFS}
\SetKwFunction{BRW}{BRW}
\RestyleAlgo{ruled}
\begin{algorithm}[ht]
\SetAlgoLined
\caption{Vortex auto-labeling algorithm in 2D CFD cases}\label{alg:autolabeling}
\Input{Nodes $\mathcal{V}$, Mass flux $\{M_{u,v}: u\in \mathcal{V}, v\in \mathcal{V}\}$, Neighborhood $\mathcal{N}(u)=\{v\in \mathcal{V}: M_{u, v}>0 \}$}
\Output{Vortex nodes: $\mathcal{V}_{vortex}$}
Build weighted directed graph $\mathcal{G=(V, E)}$ with edge weight $e_{u, v}=\frac{M_{u, v}}{\sum_{v\in \mathcal{N}(u)M_{u,v}}}$\;
initialize vortex core set $\mathcal{V}_{core}$\;
\For{each $u\in \mathcal{V}$}{
  add \DFS($\mathcal{G}, v, 0$) to $\mathcal{V}_{core}$\;
}
create working graph $\mathcal{G}_w \leftarrow \mathcal{G}$\;
\While{$\mathcal{G}_w$ changes}{
    \For{icore in $\mathcal{V}_{core}$}{
      randomly select a node $u$ from $icore$\;
      find the loop enclosing $u$: $L_u \leftarrow$\BRW($\mathcal{G}_w, u$)\;
      remove the nodes enclosed by $L_u$ from $\mathcal{G}_w$\;
    }
}
return $\mathcal{V}_{vortex} = \mathcal{G}.nodes() - \mathcal{G}_w.nodes()$;
\end{algorithm}

\begin{algorithm}[ht]
\caption{Depth first search algorithm }\label{alg:DFS}
\Input{Graph $\mathcal{G}$, starting node $v$, depth count $d$}
\Output{Vortex core nodes}
$d++1$\;
\For{ w in $\mathcal{G}.adj[v]$}{
  \uIf(\tcc*[h]{Set depth to 4 for structured mesh, 8 for unstructured mesh.}){$w$ is unvisited {\bf and} $d<depth$}{
    \If{\DFS($\mathcal{G}, w, d$)}{
      return list(w, \DFS($\mathcal{G}, w, d$))\;}
  }{\uElseIf{$w$ is visited}{
      return $w$\;
    }\Else{return None}
  }
}
\end{algorithm}

\SetKwComment{Comment}{/* }{ */}
\SetKwInOut{Input}{input}
\SetKwInOut{Output}{output}
\SetKwFunction{BRW}{BRW}
\begin{algorithm}[]
\caption{Biased random walking }\label{alg:BRW}
\Input{Graph $\mathcal{G}$, Starting point $u$}
\Output{Nodes on the loop $\mathbf{V}_{path}$ enclosing $u$}
\While(\tcc*[h]{Threshold depends on mesh size.}){step $<$ threshold}{
  add $u$ to $path$\;
  select next node $v$ in $G.adj[u]$ with possibility $\mathcal{P}(v) \propto e_{u, v}$\;
  \If{$v$ in $\mathcal{V}_{core}$}{
    return $\mathbf{V}_{path}$\;
  }
  walk to next $v: u\leftarrow v$\;
  $step++1$\;
}
\end{algorithm}

\subsection{Input selection}\label{input_selection}
\begin{figure}[!ht]\centering
     \begin{subfigure}{0.45\textwidth}\centering
         \includegraphics[scale=0.45]{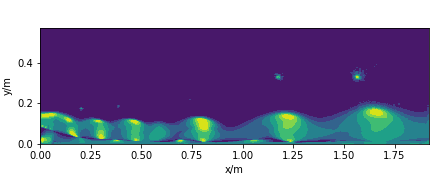}
         \caption{$Q$-criterion.}
     \end{subfigure}
     \hfill
    \begin{subfigure}{0.45\textwidth}\centering
         \includegraphics[scale=0.45]{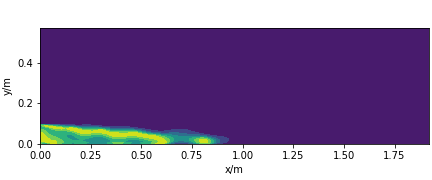}
         \caption{Turbulence intensity.}
     \end{subfigure}
     \vfill
     \begin{subfigure}{0.45\textwidth}\centering
         \includegraphics[scale=0.45]{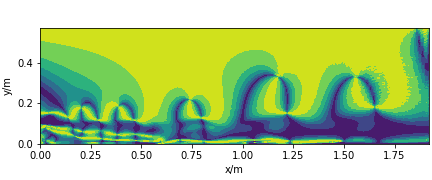}
         \caption{Deviation from shear flow.}
     \end{subfigure}
     \hfill
     \begin{subfigure}{0.45\textwidth}\centering
         \includegraphics[scale=0.45]{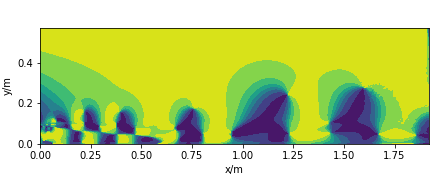}
         \caption{Pressure gradient along streamline.}
     \end{subfigure}
     \vfill
     \begin{subfigure}{0.45\textwidth}\centering
         \includegraphics[scale=0.45]{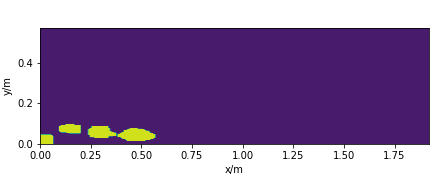}
         \caption{Ground-truth label.}
     \end{subfigure}
     \hfill
     \begin{subfigure}{0.45\textwidth}\centering
         \includegraphics[scale=0.45]{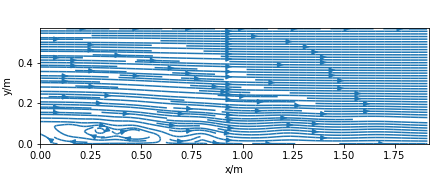}
         \caption{Streamline.}
     \end{subfigure}
\caption{The contour plot of four vortex indicators, ground-truth label and streamline of the test case at time=12.65s. }
\label{Fig.inputplot}
\end{figure}

To find out the optimal vortex indicators as the input, five input sets were tested as summarized in Table \ref{tab:FiveInputs}. The velocity field as the baseline input set and their normalized counterpart were tested. The non-dimensional $Q$ criterion and other three non-dimensional vortex indicators which have certain spatial distribution were selected. As shown in Fig. \ref{Fig.inputplot}, the turbulence intensity overlaps the vortex region. There are characteristic lines traversing through the vortex region for deviation from shear flow field and pressure gradient along streamline field. The combination of input set \#1 and input set \#4 is also tested.
\begin{table}[ht]\centering
\caption{Non-dimensional input features characterizing vortex as input for CNNs. The features are normalized following the practice in \cite{ling2015evaluation}: $ \widehat{a}=\frac{\Vert\alpha\Vert}{\Vert\alpha\Vert+\Vert \beta\Vert}$}\label{tab:FiveInputs}\centering
\begin{tabular}{c | c c}
\hline
Feature & $\alpha $ & $\beta$ \\
\hline
Normalized velocity          &   $\mathbf{U}$ & 0\\
Q-criterion                  &  $\frac{1}{2}(\Vert \mathbf{R} \Vert^2 -\Vert \mathbf{S}\Vert^2$) & $\Vert \mathbf{S}\Vert^2$ \\
Turbulence intensity & $k$ & $0.5U_iU_i + k$\\
Pressure gradient along streamline & $U_k \frac{dP}{dx_k}$ & $\sqrt{\frac{dP}{dx_j} \frac{dP}{dx_j}U_i U_i}$\\
Deviation from parallel shear flow & $\lvert U_kU_l\frac{dU_k}{dx_l}\rvert$ & $\sqrt{U_n U_n U_i \frac{dU_i}{dx_j} U_m \frac{dU_m}{dx_j} }$ \\
\hline
\end{tabular}
\end{table}

\section{Experimental study}\label{results}
\subsection{Training details}\label{training}
Firstly, five CNN models are trained on the structured mesh in order to select the optimal input set. With the selected input set, three GNN kernels are evaluated on graphs generated by AMG method from 2D BFS structured mesh and compared with CNN. The kernel size is set to 9 and $3\times3$ for GMM and SplineCNN respectively to make the their trainables comparable to CNN model, while GCN model has much less trainables than other models because of its isotropic nature. We tested GCNs with trainables at the same order as other models, however the results are not very different from the current one. For simplicity, those tests are not included here. The trainables of different models are summarized in Table \ref{tab:ParametersOfFourModles}. The GNN models are trained on graphs generated from unstructured mesh to demonstrate their adaptability to unstructured mesh. As last, in order to demonstrate the generality and limitation of the proposed method, the GNN models are trained on BFS structured and unstructured meshes and then applied to unseen cases at different Reynolds numbers simulated by different turbulence models with different meshes.
\begin{table}[h]
\caption{Summary of details of different models.}
\centering
\begin{tabular}{c c c c c}
\hline
No. & Model & Layers & Kernel size & Trainables \\
\hline\hline
1 &     CNN      & 8 & 3$\times$3 & 55633 \\
2 &     GMM      & 8 &     9      & 55921\\
3 &  SplineCNN   & 8 & 3$\times$3 & 55633\\
4 &     GCN      & 8 &     -      & 6353\\
\hline
\end{tabular}
\label{tab:ParametersOfFourModles}
\end{table}

All the trainings are trained for 100 epochs and repeated five times with different random seeds to demonstrate the stability of the model. The mean and standard deviation of the training loss and classification evaluation metrics are calculated. The Adam optimizer is used to minimize the binary cross-entropy loss function. The learning rate and weight decay rate are set to $10^{-3}$ and $10^{-5}$, respectively. The batch size is set to 5. All the GNN models are implemented in Deep Graph Library (DGL) in Pytorch framework and trained on single Intel Xeon Platinum 8260 CPU @ 2.40GHz on the cronos cluster of EDF.

\begin{figure}[!ht]
\centering
     \begin{subfigure}{0.45\textwidth}
         \centering
         \includegraphics[scale=0.35]{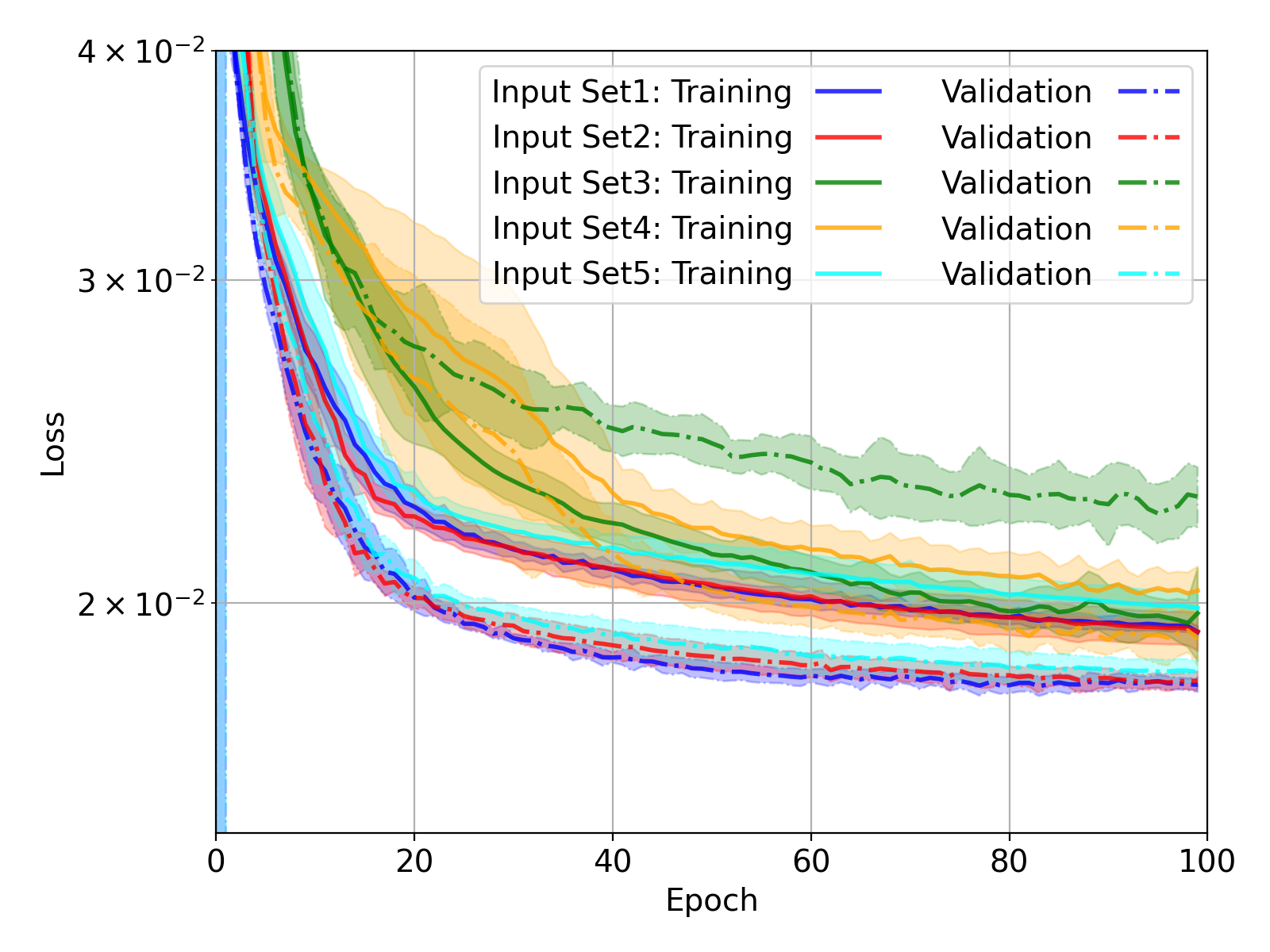}
         \caption{CNNs with five input sets.}
     \end{subfigure}
     \hfill
     \begin{subfigure}{0.45\textwidth}
         \centering
         \includegraphics[scale=0.35]{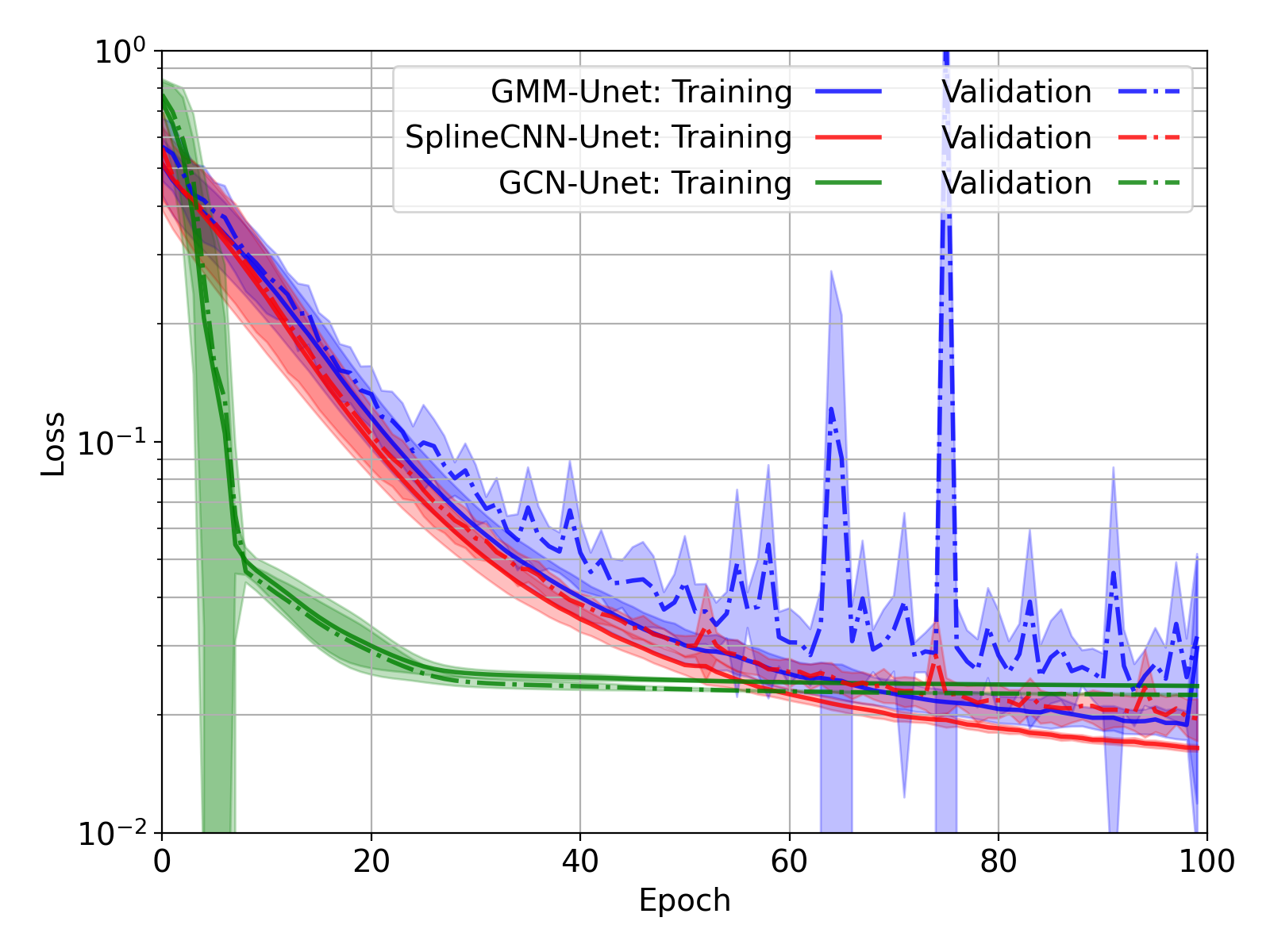}
         \caption{GNNs on AMG graphs from structured mesh.}
     \end{subfigure}
\caption{Training loss history. The curve and shaded region represent the mean and standard deviation of five trainings, respectively.}\label{Fig.loss}
\end{figure}

The training loss histories for CNNs with five input set and GNNs on AMG graphs from structured mesh are shown in Fig.\ref{Fig.loss}. CNNs give very good identifications within just 10 epochs while GNNs need more epochs. The training of all GNNs are stable except GMM which has a significant fluctuation for validation loss.

\subsection{Evaluation of input sets}
Since the dataset is a biased one where only a small portion of points belong to vortex region, therefore the following four classification metrics are used to evaluate the classification performance: accuracy, precision, recall and F1 score:
\begin{eqnarray}
Accuracy &=& \frac{TP+TN}{TP+TN+FP+FN},\\
Precision &=& \frac{TP}{TP+FP},\\
Recall &=& \frac{TP}{TP+FN},\\
F1 &=& \frac{2\times Precision\times Recall}{Precision+Recall},
\end{eqnarray}
where $TP, TN, FP, FN$ are the numbers of true positives, true negatives, false positives and false negatives, respectively. The performances of five input sets are evaluated in the region behind the step ($x>0$) with CNN-Unet algorithm and summarized in Table \ref{tab:FiveInputsPerformance}. The highest and lowest values of each of four classification evaluation metrics among all input sets are labeled in green and red color, respectively. It is obvious that input sets \#1 and \#2 are generally better than other input sets in terms of accuracy, precision and F1 score, while the input sets \#4 has the worst performance. Input set \#2 outperforms other input sets on training time per epoch by a large margin. Input set \#3 (Q-criterion) has the highest standard deviations for all metrics and it also needs the longest training time. Compared with input sets \#1 and \#2, additional inputs in Input set \#5 do not bring performance improvement but the increased training time.

\setlength\tabcolsep{6.0pt}
\begin{table}[h]
\caption{Classification performance of CNNs trained on five different input sets. (\textcolor{green}{Green: best value}; \textcolor{red}{Red: worst value}.)}\centering
\begin{tabular}{c| c c c c c}
\hline
No. & Accuracy & Precision & Recall & F1 score & Time/epoch\\
\hline
1 & \textcolor{green}{88.03}±0.16 & 90.86\textcolor{green}{±0.48}& 61.73±0.81 & \textcolor{green}{73.51}±0.50 &  2.63±0.79s\\
2 & 88.02±0.27 & \textcolor{green}{91.82}±0.49 & 60.92±0.94 & 73.24±0.74 &  \textcolor{green}{1.85}±0.13s\\
3 & 87.83±\textcolor{red}{1.10} & \textcolor{red}{89.38±3.97} & \textcolor{green}{62.74}±\textcolor{red}{7.78} & 73.22±\textcolor{red}{4.32} &  \textcolor{red}{4.43}±0.34s\\
4 & \textcolor{red}{87.53}±0.31 & 90.42±0.86 & 60.03±1.23 & \textcolor{red}{72.15}±0.89 &  2.42±0.27s\\
5 & 87.64±\textcolor{green}{0.12} & 91.44±0.49 & \textcolor{red}{59.68}±\textcolor{green}{0.61} & 72.22±\textcolor{green}{0.37} &  3.32±1.63s\\
\hline
\end{tabular}\label{tab:FiveInputsPerformance}
\end{table}

\begin{figure}[!ht]
\centering
\includegraphics[scale=0.4]{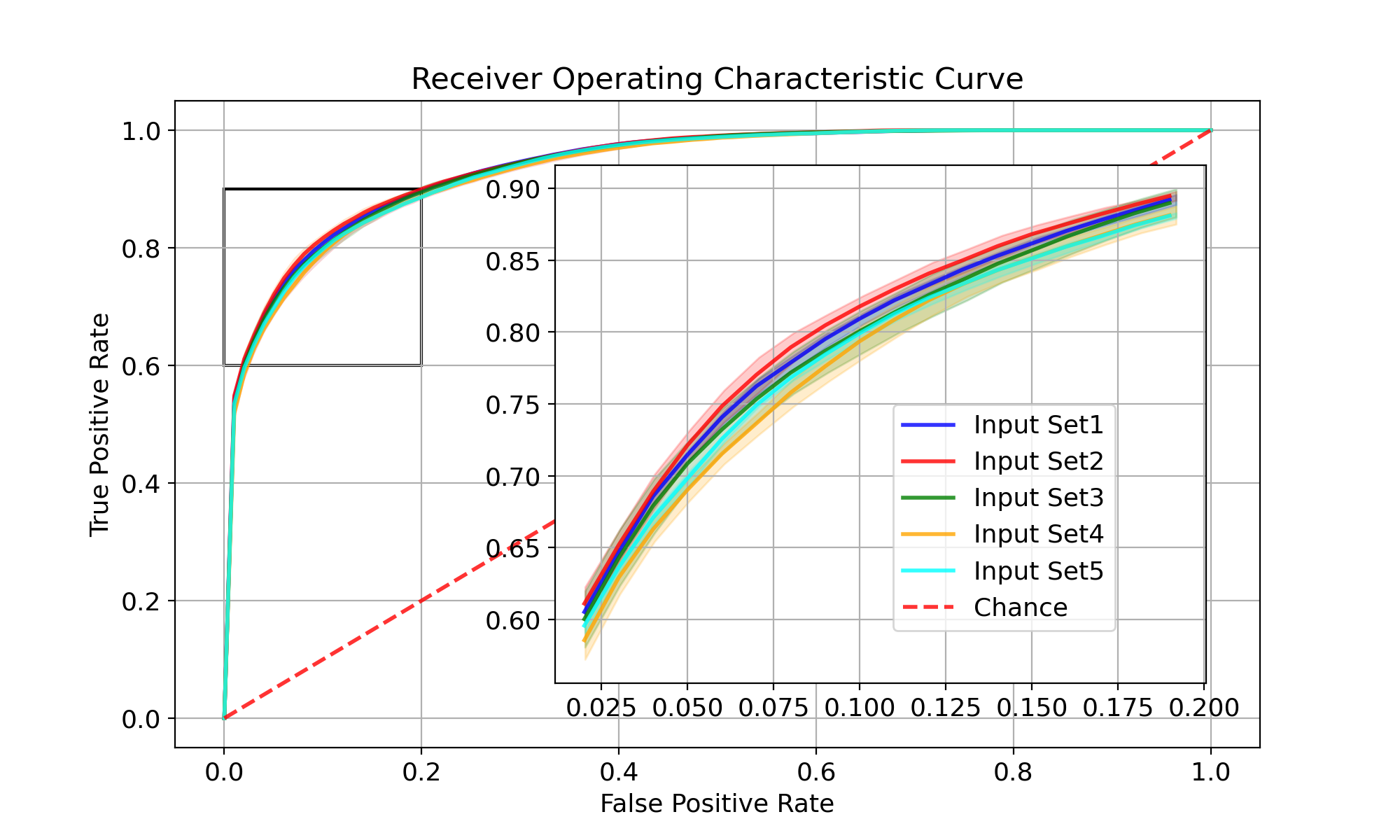}
\caption{Receiver operating characteristic curves of CNN-Unet with different input sets. (The curve and shaded region represent the mean and standard deviation of five trainings respectively.)}
\label{Fig.roc_input_selection}
\end{figure}

A receiver operating characteristic (ROC) curve \cite{fawcett2006introduction} is a graphical plot that illustrates the diagnostic ability of a binary classifier system as its discrimination threshold varies. The ROC curve is created by plotting the true positive rate (TPR) against the false positive rate (FPR) at various threshold settings. The ideal prediction model should yield a point in the upper left corner or coordinate (0,1) in the ROC space, representing no false negatives and no false positives. As shown in Fig. \ref{Fig.roc_input_selection}, the ROC curves of all input sets are very close and input set \#2 is slightly better than the others.

\begin{figure}[!ht]
\centering
\includegraphics[scale=0.2]{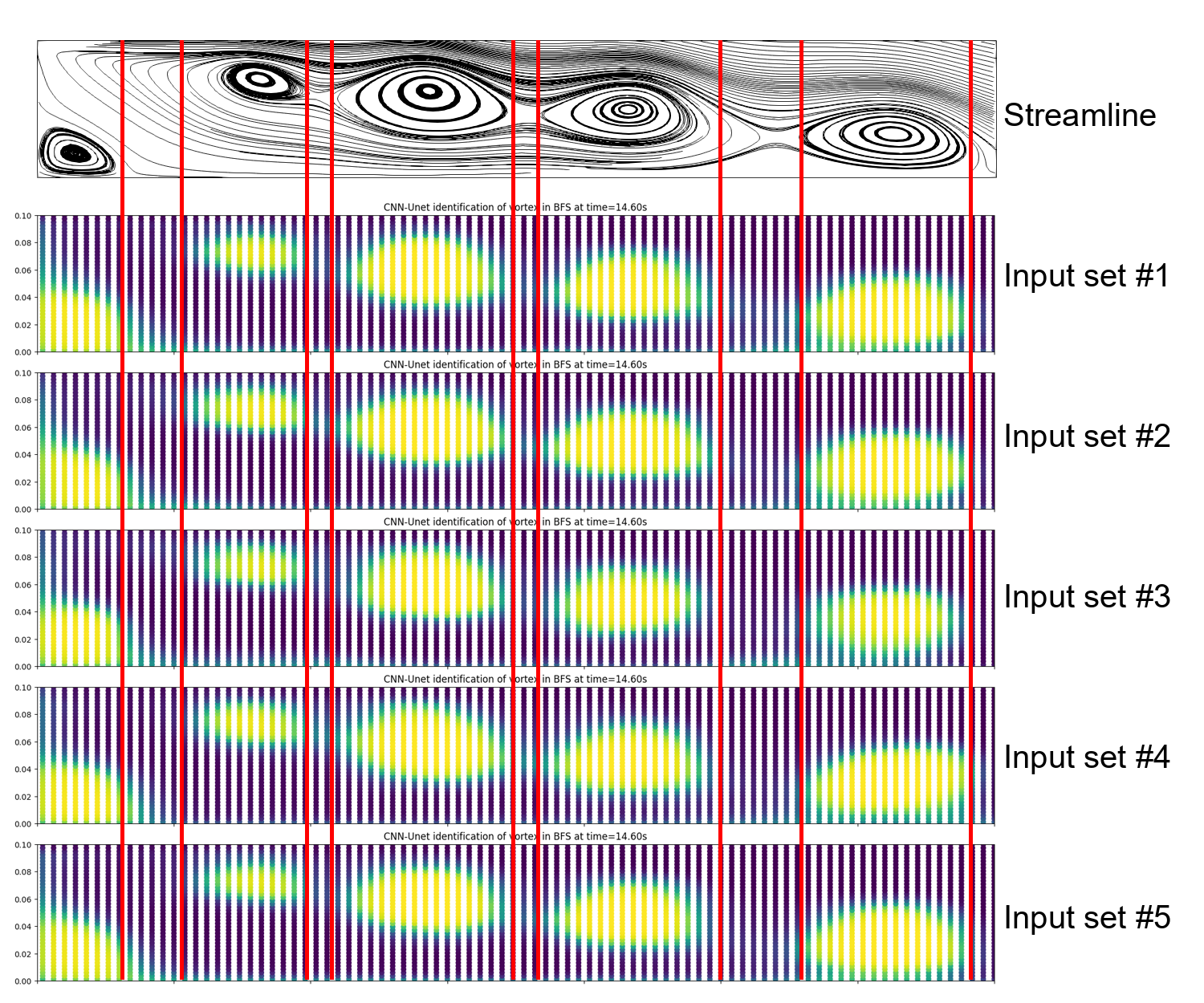}
\caption{Vortexes at time=14.60s identified by CNN-Unet algorithm with different input sets. From top to bottom: streamline, input sets from \#1 to \#5.}
\label{Fig.FiveInputSetsclassification}
\end{figure}

As shown in Fig. \ref{Fig.FiveInputSetsclassification}, although all input sets can be used to precisely identify the vortex position, those identified with input sets \#2 have the shapes closest to the real vortexes. Since with input set \#2, CNN-Unet achieves a good performance on four classification evaluation metrics, has the best ROC curve and best identified vortexes shape, and shortest training time per epoch, we continue to compare different GNN-based models based on input set \#2 - normalized velocity field.

\subsection{Evaluation of GNN kernels}\label{GNNKernelsEvuation}
We evaluate the performance of three GNN-based kernels on graphs generated by AMG from structured mesh by comparing them with training No. 2 of CNN models in Table \ref{tab:FiveInputsPerformance} with the same input set - normalized velocity field. 

\noindent\textbf{Classification performance.}
The classification performance of four models is summarized in Table \ref{tab:GNNsOnAMG}. GMM kernel performance is very close to that of CNN kernel. SplineCNN kernel outperforms CNN kernel with higher mean values of all classification evaluation metrics except for precision and its standard deviations are smallest among four GNN kernels although higher than CNN. Among all models, GCN model has the worst classification performance, which is expected. 

Regarding the computational efficiency, it should be pointed out that GCN has disproportionately longer training time compared to its significantly smaller trainables, which suggests that it struggles to back-propagate the gradients since it cannot identify the features with spatial distribution. Except GCN, CNN training time is one order lower than those of GNN-based models. The training time of SplineCNN is the highest and 1.5 times longer than that of GMM.

\setlength\tabcolsep{3.0pt}
\begin{table}[h]
\fontsize{8}{12}\selectfont 
\caption{Comparison of classification performances of GNN kernels on graphs generated by AMG with CNN. (Among three GNN-based models: \textcolor{green}{Green: best value}; \textcolor{red}{Red: worst value}.)}
\centering
\begin{tabular}{c c c c c c c}
\hline
Model & Accuracy & Precision & Recall & F1 score & Training time/epoch & Inference time/case \\
\hline\hline
     CNN      & 88.02±0.27 & 91.82±0.49 & 60.92±0.94 & 73.24±0.74 &  1.85±0.13s & 0.008s\\
     GMM      & 88.13±\textcolor{green}{0.14} & 87.95±1.41 & 64.81±\textcolor{green}{1.69} & 74.59±\textcolor{green}{0.65} & 44.14±0.11s & 0.291s \\
  SplineCNN   & \textcolor{green}{89.28}±\textcolor{red}{0.37} & \textcolor{green}{91.70±0.63} & \textcolor{green}{66.17}±2.03 & \textcolor{green}{76.84}±1.15 & \textcolor{red}{67.59}±0.16s &  0.553s\\
     GCN      & \textcolor{red}{76.47}±0.23 & \textcolor{red}{62.17±1.42} & \textcolor{red}{32.29±2.12} & \textcolor{red}{42.43±1.71} & \textcolor{green}{10.94}±0.03s & 0.097s\\
\hline
\end{tabular}\label{tab:GNNsOnAMG}
\end{table}

\begin{figure}[!ht]
    \centering
    \begin{subfigure}{0.48\textwidth}
        \includegraphics[scale=0.17]{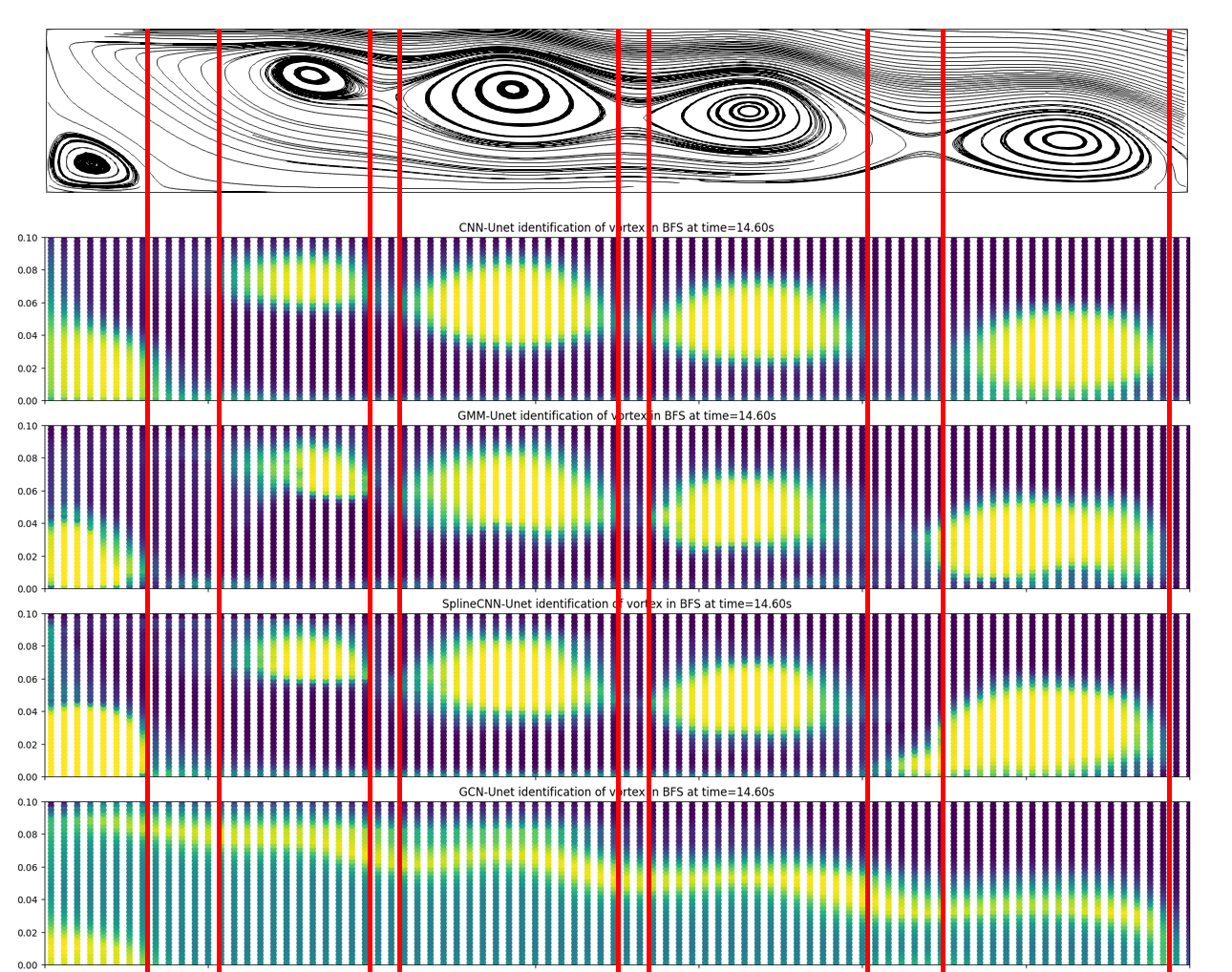}
        \caption{Identified vortex at time=14.60s. From top to bottom: streamline, CNN, GMM, SplineCNN and GCN.}\label{Fig.fourmodelclassification}
    \end{subfigure}
    \hfill
    \begin{subfigure}{0.45\textwidth}
        \includegraphics[scale=0.35]{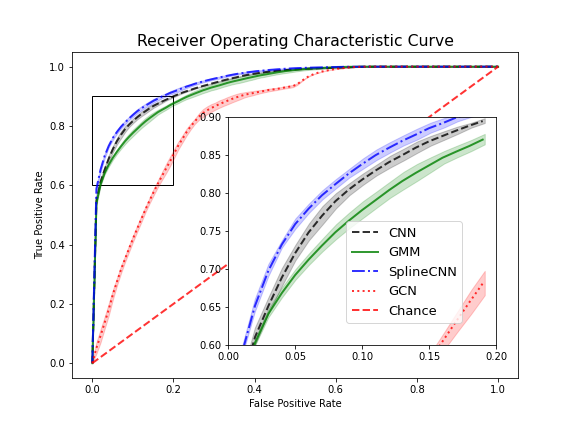}
        \caption{Receiver operating characteristic curves. (The curve and shaded region represent the mean and standard deviation of five trainings respectively.) }\label{Fig.ROC_4Models}
    \end{subfigure}
    \caption{Comparison of GNN kernels against CNN kernel.}
\end{figure}

\noindent\textbf{Vortex morphology.} As shown in Fig. \ref{Fig.fourmodelclassification}, the shapes and positions of the vortexes identified by CNN and SplineCNN kernels are very close to each other, both are very clear and accurate. By comparison, the vortexes' boundary identified by GMM kernel is diffused. GCN kernel fails to identify neither the vortex shape nor position. 

\noindent\textbf{ROC curve.} As shown in Fig. \ref{Fig.ROC_4Models}, among all models, GCN kernel has the worst ROC and SplineCNN kernel has the ROC curve closest to the left upper corner which means it's classification can obtain the highest positive rate and the lowest false positive rate. The ROC curves of CNN and SplineCNN kernel are very close to CNN's ROC  and slightly better than that of GMM. 

\subsection{Adaptability to unstructured mesh}
To demonstrate their adaptability to unstructured meshes, the three GNN kernels are trained on graphs generated by AMG method from unstructured mesh, whose edge and node information is summarized in Table \ref{table.GraphsDetails}. As shown in Table \ref{tab:GNNsOnAMG_unstruct}, SplineCNN outperforms GMM on four classification evaluation metrics, higher mean values and smaller standard deviations, especially for recall and F1 score whose mean values exceed those of GMM by a considerable margin. However, the better classification performance of SplineCNN is also accompanied by considerable computational overhead, about $66.6\%$ higher than GMM which is a disadvantage dealing with 3D CFD cases with much more mesh cells in the future. Three of five GCN trainings did not converge resulting in the Nan±Nans in the table.

\setlength\tabcolsep{3.0pt}
\begin{table}[h]
\fontsize{8}{12}\selectfont 
\caption{Classification performance of GNN kernels on graphs generated by AMG from unstructured mesh. (Between GMM and SplineCNN: \textcolor{green}{Green: best value}; \textcolor{red}{Red: worst value}.)}
\centering
\begin{tabular}{c c c c c c c}
\hline
   Kernel   & Accuracy   & Precision  & Recall     & F1 score   & Training time/epoch & Inference time/case \\
\hline\hline
    GMM      & \textcolor{red}{92.76±0.27} & \textcolor{red}{81.00±1.58} & \textcolor{red}{67.27±3.22} & \textcolor{red}{73.43±1.55} & \textcolor{green}{104.05}±0.34s & 0.755s \\
 SplineCNN   &  \textcolor{green}{93.54±0.26} &  \textcolor{green}{82.21±1.11} &  \textcolor{green}{72.41±2.42} &  \textcolor{green}{76.97±1.23} &\textcolor{red}{173.32}±0.20s & 1.530s \\
    GCN      & 85.60±0.68 &   Nan±Nan  & 9.73±12.14 &   Nan±Nan  & 19.72±0.11s & 0.223s\\
\hline
\end{tabular}
\label{tab:GNNsOnAMG_unstruct}
\end{table}

\begin{figure}[!ht]
    \centering
    \begin{subfigure}{0.5\textwidth}
        \includegraphics[scale=0.12]{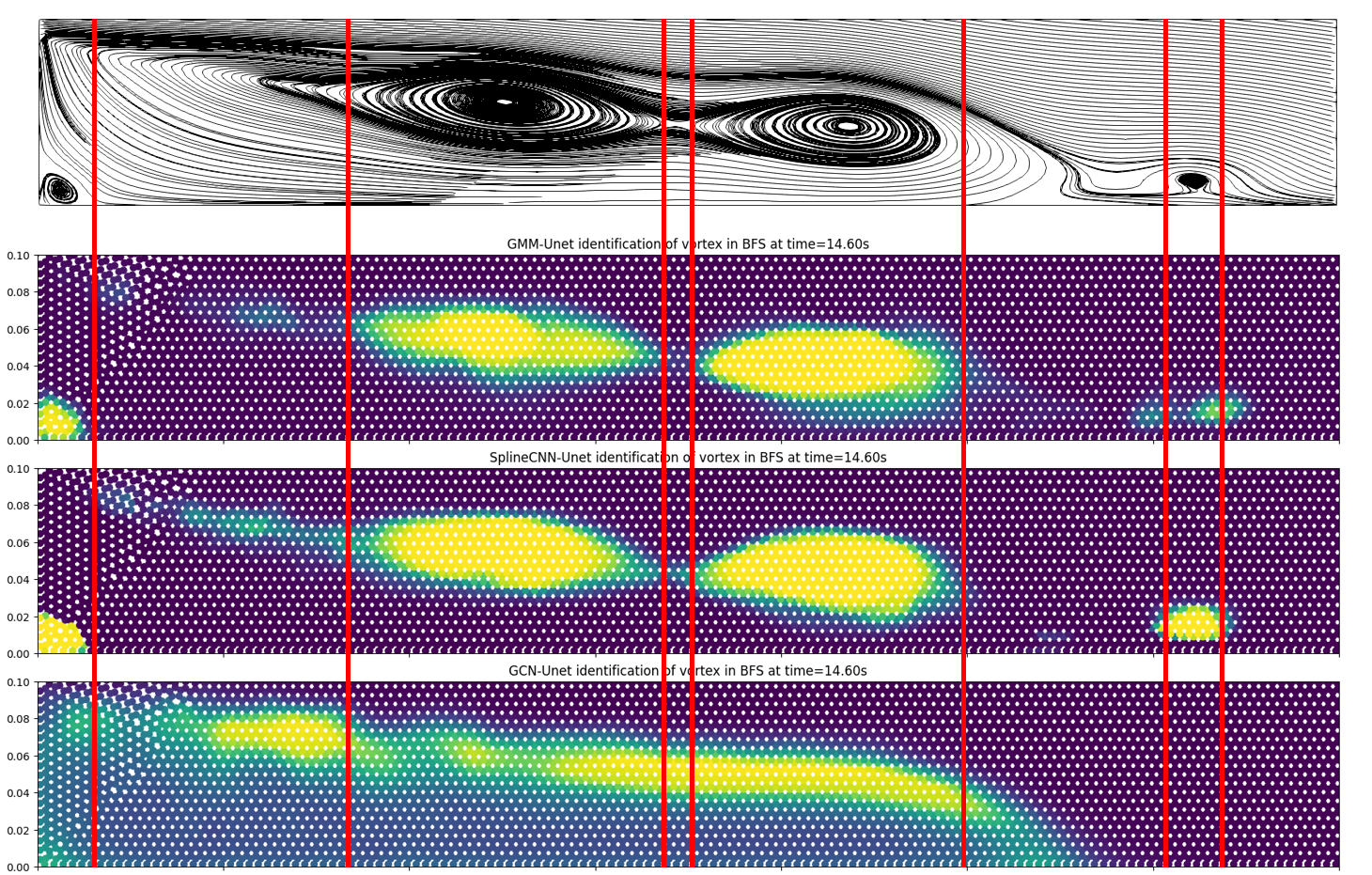}
        \caption{Identified vortex.}\label{fig:GNNsAMGUnstructClassification}
    \end{subfigure}
    \hfill
    \begin{subfigure}{0.45\textwidth}
        \includegraphics[scale=0.3]{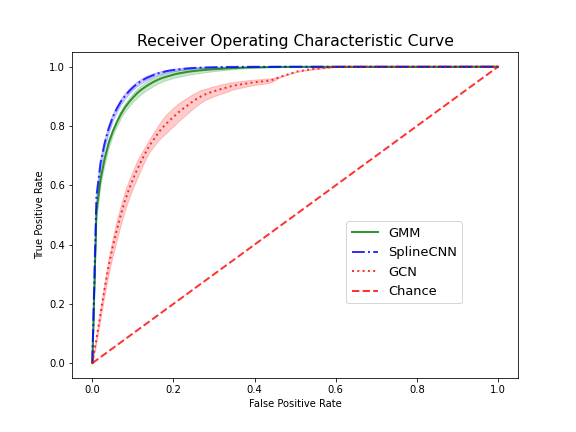}
        \caption{Receiver operating characteristic curves. (The curve and shaded region represent the mean and standard deviation of five trainings respectively.) }\label{fig:rocGNNsAMGUnstruct}
    \end{subfigure}
    \caption{GNNs on unstructured mesh.}
\end{figure}

As shown in Fig. \ref{fig:GNNsAMGUnstructClassification}, GMM well identifies the shape of the small vortexes and fails on elongated vortexes while SplineCNN does the opposite. GCN can neither identify the vortex shape nor the position. As shown in Fig. \ref{fig:rocGNNsAMGUnstruct}, the ROC curve of SplineCNN is very close to but slightly better than that of GMM and both ROC curves of two anisotropic kernels are far better than that of isotropic kernel - GCN.

\subsection{Generality analysis}
\begin{figure}[!ht]
    \begin{subfigure}{0.5\textwidth}
        \centering
        \includegraphics[scale=0.2]{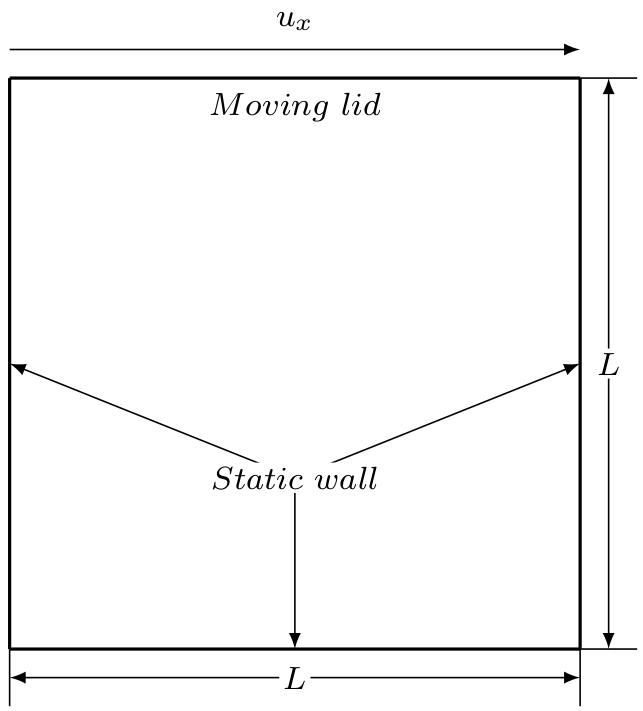}
        \caption{Lid-driven cavity.}\label{fig.Cavity}
    \end{subfigure}
    \hfill
    \begin{subfigure}{0.5\textwidth}
        \centering
        \includegraphics[scale=0.2]{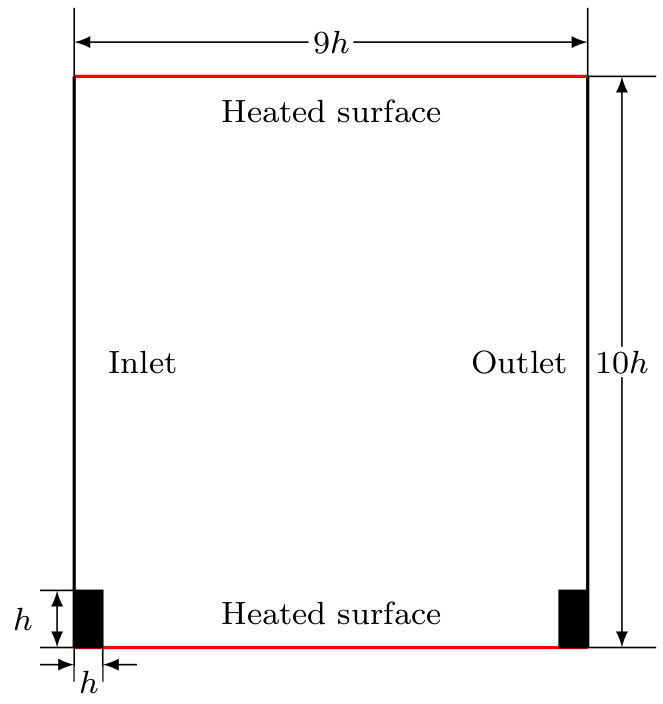}
        \caption{RIBS.}\label{fig.RIBS}
    \end{subfigure}
    \vfill
    \begin{subfigure}{1\textwidth}
        \centering
        \includegraphics[scale=0.2]{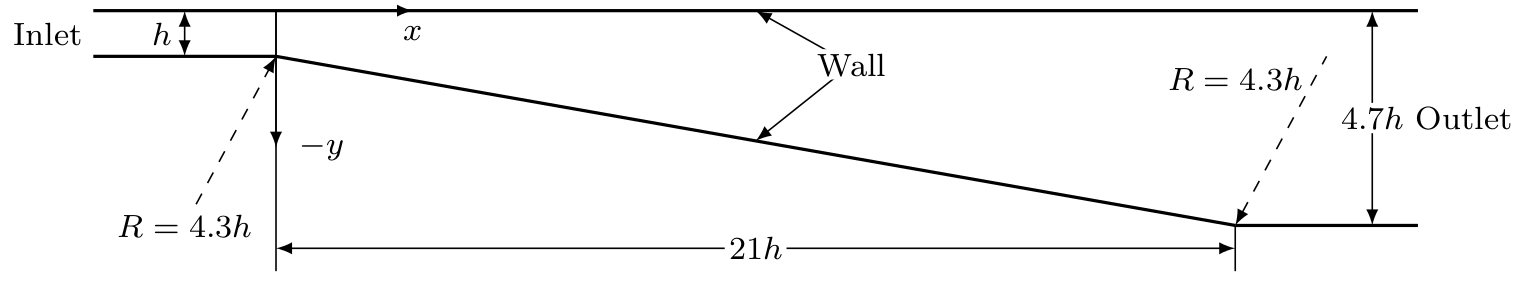}
        \caption{Diffuser.}\label{fig.Diffuser}
    \end{subfigure}
    \caption{Configurations of three cases.}\label{fig.3caseConfiguration}
\end{figure}

To further evaluate the generality of the proposed approach on detecting vortexes with respect to different meshes in terms of mesh type, mesh density and mesh aspect ratio for structured mesh, different turbulence models and different Reynolds numbers, the GMM-Unet and SplineCNN-Unet models are trained again on the dataset formed by the BFS results obtained using both BFS structured and unstructured meshes and then tested on the three of code\_saturne validation cases: lid-driven cavity flow, heat transfer in a cooling channel with periodic ribs (RIBS)\cite{rau1998effect, arts2007experimental}, asymmetric plane diffuser flow\cite{buice1997experimental, obi1993experimental}. The main simulation details of these cases are summarized in Table \ref{tab.foucases}. The configuration of three cases are shown in Fig. \ref{fig.3caseConfiguration}.

\noindent\textbf{Lid-driven cavity.} 
The lid-driven cavity flow configuration is shown in Fig. \ref{fig.Cavity}. The top lid moves towards right direction and other walls are static. The no-slip conditions are applied on the walls. The Reynolds number is 5000. The $k-\omega\ SST$ turbulence model was used to simulated the cavity flow on the finest mesh which contains $N_x\times N_y =300\times 300$ cells. The velocity field is then interpolated to coarser meshes of different refinement levels from 2500 to 40000 cells of the same aspect ratio $AR = 1$, and two meshes of different aspect ratios from 1 to 9 with the cell number around 2500. The aspect ratio is defined as the ratio of the cell's length on the x direction to its width on the y direction.

\noindent\textbf{RIBS.} 
The RIBS configuration is shown in Fig. \ref{fig.RIBS}. Air flows from the left to the right, at the atmospheric pressure and temperature. The left and right boundaries are set to be periodic. The top and bottom walls are heated while the two ribs are not. The Reynolds number and Prandtl number are 30000 and 0.71, respectively. The RIBS case was simulated using three turbulence models with both structured and unstructured meshes.

\noindent\textbf{Diffuser.} 
The 2D flow inside a planar asymmetric diffuser, as shown in Fig. \ref{fig.Diffuser}, is simulated using $k-\omega\ SST$ turbulence model with fully-developed turbulent inlet at $Re=18000$ based on the bulk inlet velocity and the inlet channel height with two types of meshes: high-Reynolds (HR) mesh and low-Reynolds (LR) mesh. 
\begin{table}[h]
\centering
\caption{Simulations details of four cases.}
\centering
\begin{tabular}{c | c | c | c c c}
\hline
\multirow{2}{*}{Case} & \multirow{2}{*}{$Re$} & \multirow{2}{*}{Turbulence model} & \multicolumn{3}{c}{Mesh}\\
\cline{4-6}
 & & & Type & AR & $No.\ cells_{(N_x\times N_y)}$ \\
\hline
\multirow{2}{*}{BFS} & \multirow{2}{*}{5100} & \multirow{2}{*}{$R_{ij}-\epsilon\ SSG$} & Structured & 3.2 & 74400 \\
& & & Unstructured & - & 131118 \\
\hline
\multirow{7}{*}{Cavity} & \multirow{7}{*}{5000} & \multirow{7}{*}{$k-\omega\ SST$} & Structured & 1 & $90000_{300\times300}$ \\ 
& & & Structured & 1 & $40000_{200\times200}$ \\ 
& & & Structured & 1 & $10000_{100\times100}$ \\ 
& & & Structured & 1 & $2500_{50\times50}$ \\
& & & Structured & 3 & $2494_{29\times86}$ \\ 
& & & Structured & 6 & $2500_{20\times125}$ \\ 
& & & Structured & 9 & $2499_{14\times147}$ \\ 
\hline
\multirow{3}{*}{RIBS} & \multirow{3}{*}{30000} & $k-\epsilon\ LP$  & \multirow{3}{*}{\Centerstack{Structured \\ Unstructured}}& \multirow{3}{*}{\Centerstack{3\\ -}}& \multirow{3}{*}{\Centerstack{18296\\7735}} \\
& & $k-\omega\ SST$ & & & \\
& & $R_{ij}-\epsilon\ SSG$ & & &\\
\hline
\multirow{2}{*}{Diffuser} & \multirow{2}{*}{18000} & \multirow{2}{*}{$k-\omega\ SST$} & HR structured & $0.3\sim 47.7$ & $21648_{328\times 66}$\\
 & & & LR structured & $0.3\sim 119.8$ & {$31488_{328\times 96}$} \\
\hline
\end{tabular}
\label{tab.foucases}
\end{table}

\begin{figure}[!ht]
\centering
\includegraphics[scale=0.4]{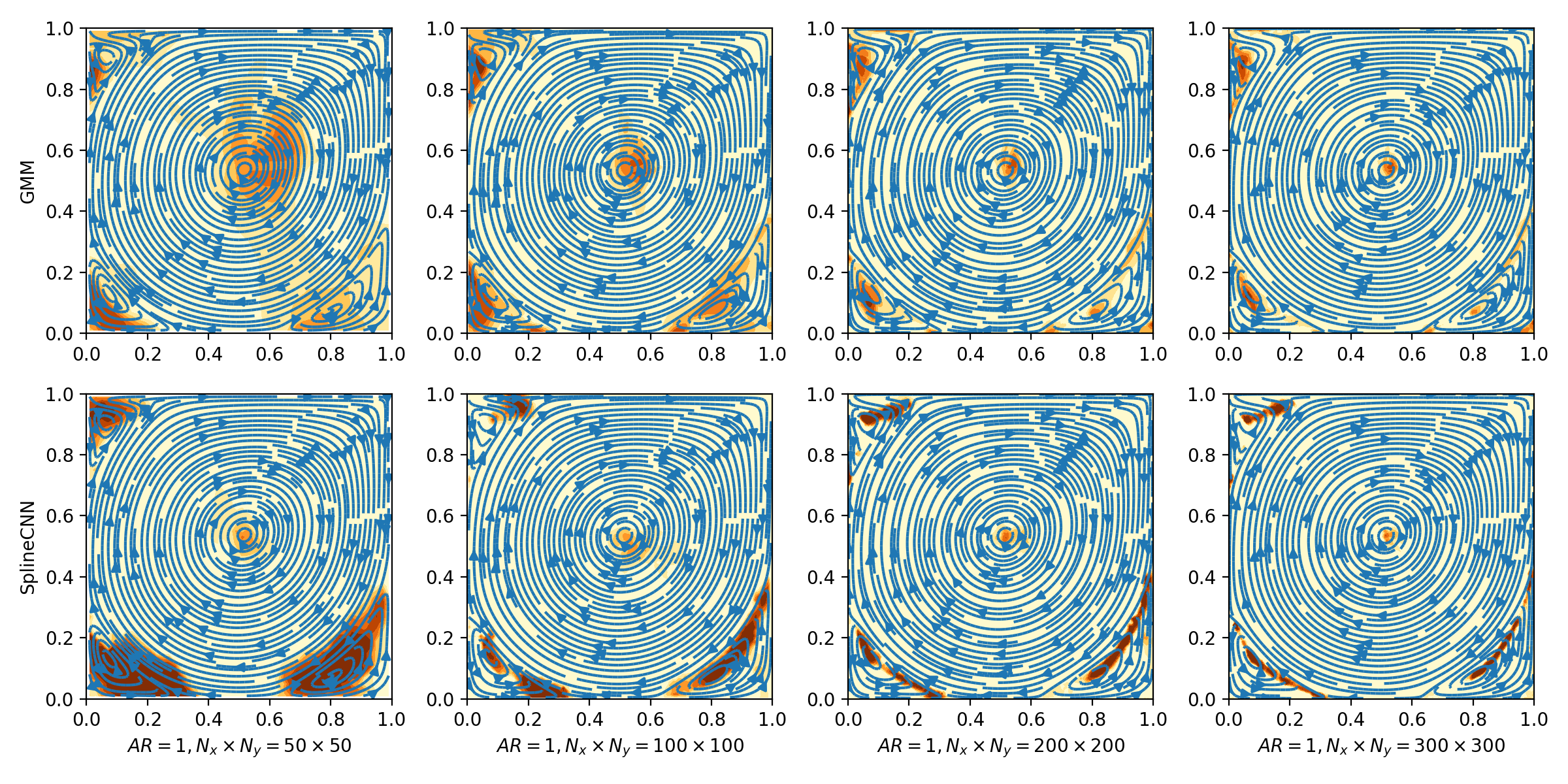}
\caption{Identifications of vortexes of GMM-Unet and SplineCNN-Unet on lid-driven cavity meshes of different refinement levels.}
\label{Fig.Meshdensity}
\end{figure}

\bigskip\noindent\textbf{Mesh density influence.} 
As show in Fig. \ref{Fig.Meshdensity}, the vortex regions identified by both GMM-Unet and SplineCNN-Unet in the coarsest mesh are the largest in the coarsest mesh. The identified regions become smaller as the mesh refinement level increases. The capability of recognizing a certain pattern of all the pure convolution-based machine learning models is limited to the size of effective receptive field (ERF). A trained model fails to correlate two points separated by a distance larger than the ERF which is determined by both the hyper-parameters of the model and the dataset. As indicated by one typical snapshot of streamline plot in Fig. \ref{Fig.bfs}, for the BFS structured mesh, $100\times 40$ cells are distributed in the vortex region where normally four to five vortexes exist. Thus, a single vortex in the training dataset covers no more than 25 cells on one specific direction. As a result, these two models trained on this dataset can only detect the vortexes of comparable sizes in terms of how many mesh cells they span.

\begin{figure}[!ht]
\centering
\includegraphics[scale=0.4]{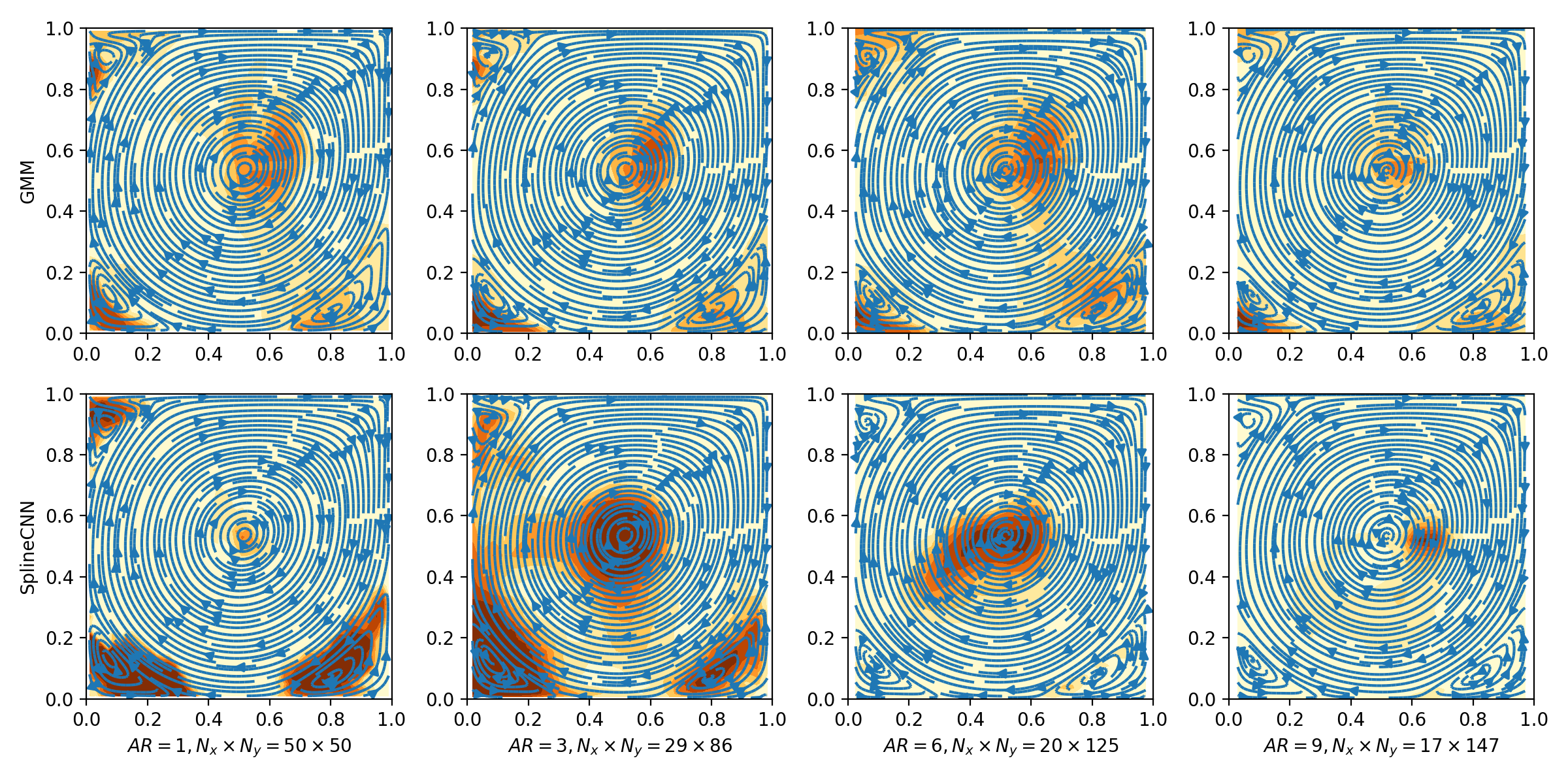}
\caption{Identifications of vortexes of GMM-Unet and SplineCNN-Unet on lid-driven cavity meshes of different aspect ratios.}
\label{Fig.MeshAR}
\end{figure}

\noindent\textbf{Mesh aspect ratio influence.} 
The aspect ratio of the BFS structured mesh included in the training dataset is $AR=3.2$. As a result, the two models trained on this dataset well identified the vortexes on the mesh of $AR=3$ as shown in Fig. \ref{Fig.MeshAR}. As the aspect ratio deviates from 3.2, their identification performances deteriorate which is more evident for SplineCNN-Unet. The SplineCNN-Unet model successfully identified three secondary vortexes in the corners on the mesh of $AR=1$, all vortexes on the mesh of $AR=3$, and the primary vortex center on the mesh of $AR=6$, but failed on the mesh of $AR=9$. Compared to SplineCNN-Unet, the GMM-Unet model identified more or less the same vortex region on the four meshes and thus has a better generality to the variation of the mesh aspect ratio. However, the vortex regions identified by the GMM-Unet model do not cover the vortex center and have diffuse boundaries compared with the SplineCNN-Unet.

\begin{figure}[!ht]
    \begin{subfigure}{1\textwidth}
        \centering
        \includegraphics[scale=0.4]{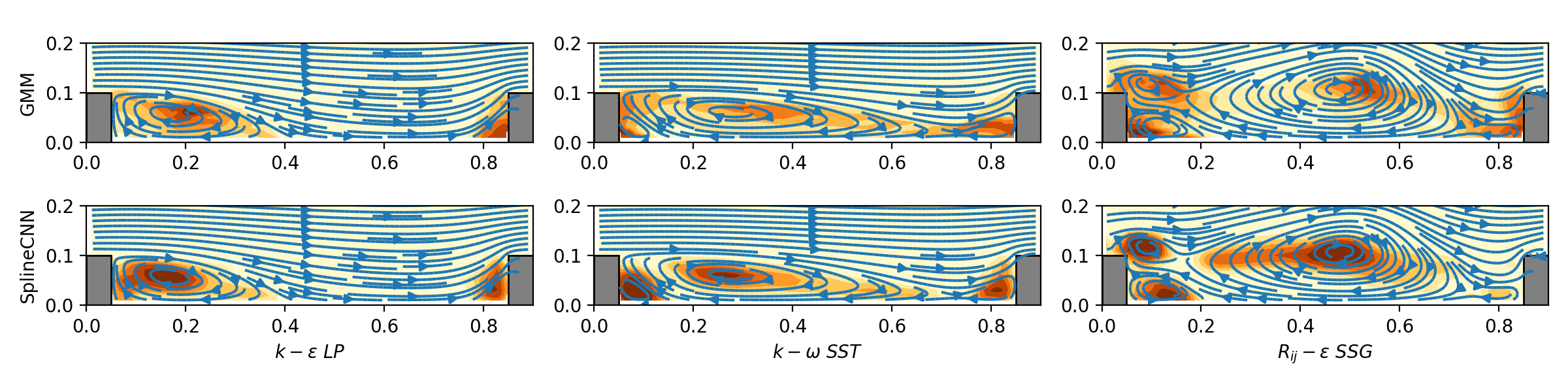}
        \caption{Structured mesh.}\label{fig.RibsStruct}
    \end{subfigure}
    \vfill
    \begin{subfigure}{1\textwidth}
        \centering
        \includegraphics[scale=0.4]{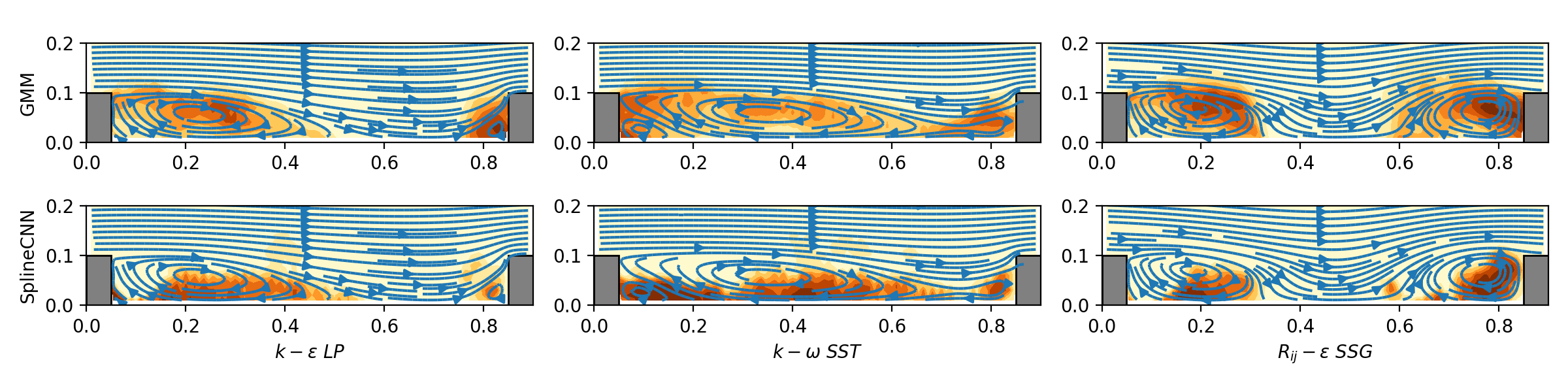}
        \caption{Unstructured mesh}\label{fig.RibsUnstruct}
    \end{subfigure}
    \caption{Identifications of vortexes of GMM-Unet and SplineCNN-Unet on RIBS structured and unstructured meshes.}\label{fig.turb_meshtype}
\end{figure}

\noindent\textbf{Mesh type influence.}
The mixture of both structured and unstructured meshes in the dataset poses no problem to the training. As shown in Fig. \ref{fig.turb_meshtype}, while the SplineCNN-Unet model better identified the vortex center and shape on the structured mesh compared with the GMM-Unet model, but degraded more on the unstructured mesh where it only identified the lower half of the vortexes. The GMM-Unet model once again has better generality to the mesh type.

\noindent\textbf{Turbulence models influence.}
To test the sensitivity of the proposed approach to the turbulence models, three commonly used models, $k-\epsilon$ linear production, $k-\omega\ SST$ and $R_{ij}-\epsilon\ SSG$, were selected because we intend to detect the vortexes generated by RANS turbulence models. As shown in Fig. \ref{fig.turb_meshtype}, the turbulence models have no visible influence on the identification performance of the proposed approach. The robustness of our models to different turbulence models is explainable since they identify the vortexes based on the topological distribution of the velocity field which is universal among different turbulence models.

\begin{figure}[!ht]
    \begin{subfigure}{1\textwidth}
        \centering
        \includegraphics[scale=0.25]{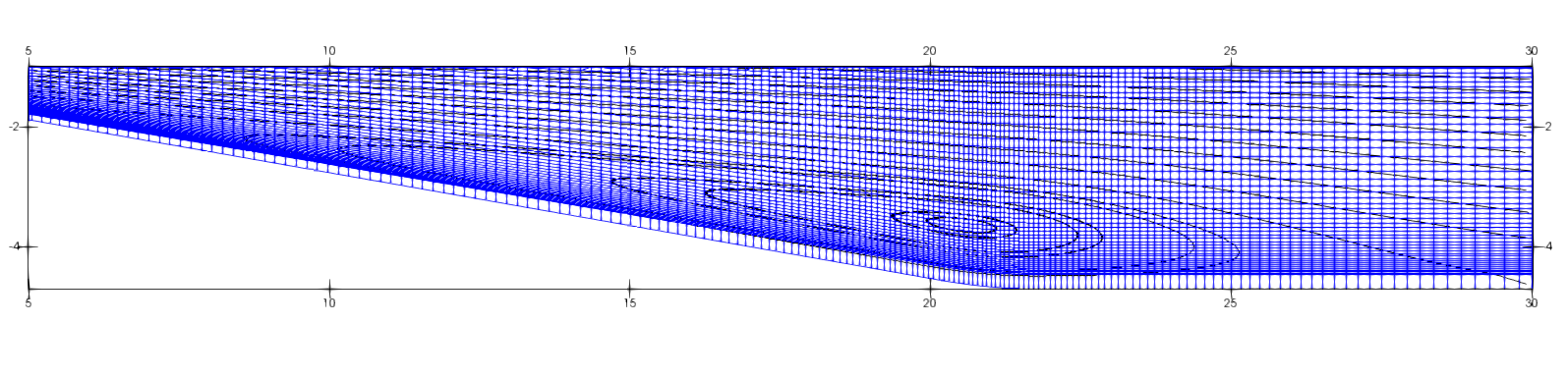}
        \caption{High-Reynolds mesh.}\label{fig.DiffuserHR}
    \end{subfigure}
    \vfill
    \begin{subfigure}{1\textwidth}
        \centering
        \includegraphics[scale=0.25]{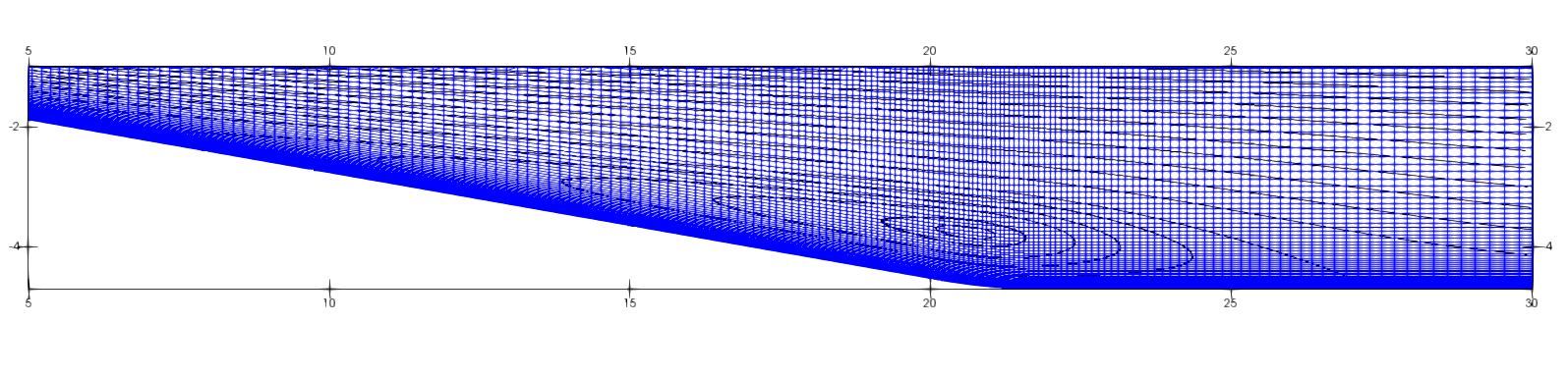}
        \caption{Low-Reynolds mesh}\label{fig.DiffuserLR}
    \end{subfigure}
    \caption{Two diffuser meshes.}\label{fig.DiffuserMesh}
\end{figure}

\begin{figure}[!ht]
\centering
\includegraphics[scale=0.45]{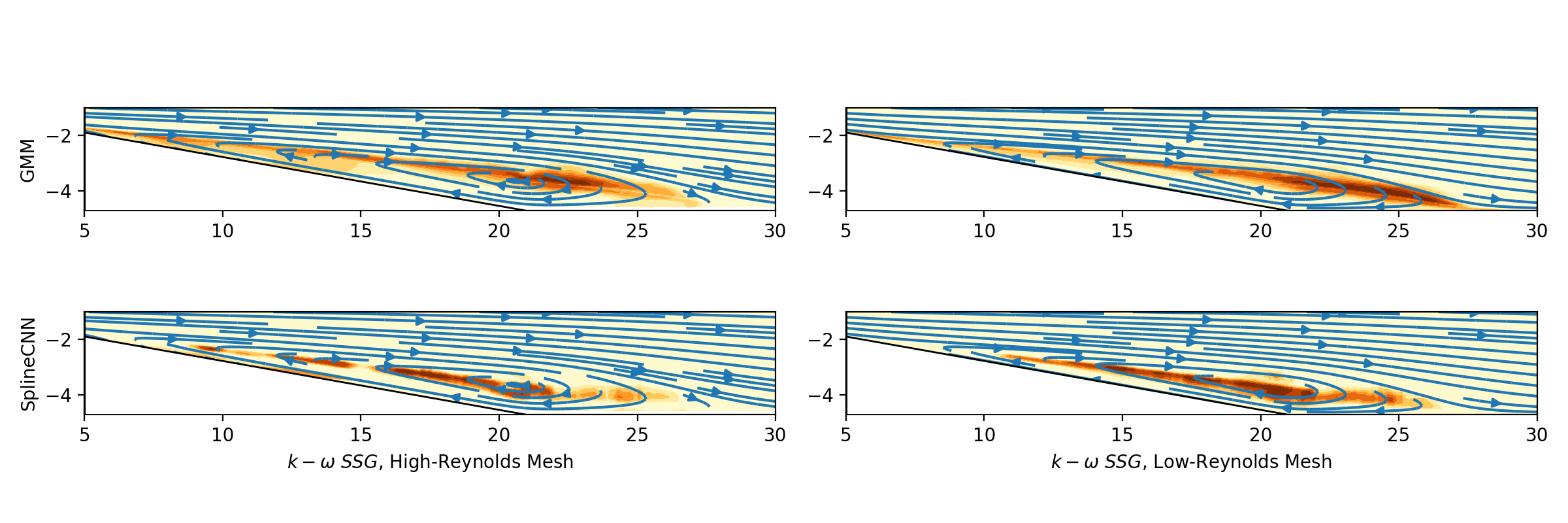}
\caption{Identifications of vortexes of GMM-Unet and SplineCNN-Unet on two diffuser meshes.}
\label{Fig.Diffuser}
\end{figure}

\noindent\textbf{Mesh size scaling influence.} 
As shown in Fig. \ref{fig.DiffuserMesh}, the cell size along the wall normal direction in the low-Reynolds diffuser mesh increases continuously, while in high-Reynolds diffuser mesh, the cell size from the first to the second layer perpendicular to the wall decreases abruptly. As shown in Fig. \ref{Fig.Diffuser}, both GMM-Unet and SplineCNN-Unet models can capture the vortexes on two meshes. On the high-Reynolds mesh, the identified vortex regions by both models are non-connected while those on the low-Reynolds mesh are closer to the real vortex topology. Therefore the proposed approach is quite robust to the mesh size scaling but sensitive to the abrupt scaling jump.

\section{Conclusion}\label{conclusion}
In this paper, we proposed a machine learning approach for identifying vortexes in unstructured mesh-based CFD results using a Graph Neural Network with U-Net architecture, which includes constructing graphs from CFD meshes, the generation of a hierarchy of graphs using algebraic multigrid method and the labeling of the dataset with the biased random walking algorithm. We demonstrated that the current machine learning approach combined with either Gaussian mixture model or SplineCNN as convolution kernel can achieve similar identification performance to that of traditional CNN while directly accepting embedded data on unstructured meshes. While the SplineCNN kernels achieves better vortex identification performance on cases similar to those included in the dataset, it degrades more significantly than GMM on unseen cases and meshes. Compared with other influence factors, such as different turbulence models and different Reynolds numbers, the mesh density has the biggest influence on the vortex identification of the GNN models. The mesh sensitivity analysis shows that the trained GNN models can only identify vortexes of the scale similar to those included in the dataset. It means that the region of the same vortex identified by the GNN model shrinks in denser meshes. This problem can be partially alleviated by including larger vortexes in the dataset but can not be fundamentally resolved. In the future, we will try to solve this problem by introducing other modules to the model architecture to exploit the invariant vortex features in much deeper graph hierarchies. The next step is to extend our approach to 3D cases and to other flow phenomena. Since we are working on graphs, this approach is extensible to 3D case without extra effort on the GNN model side but requires much efforts on generating a comprehensive 3D dataset. 

\section*{Acknowledgements}
This work has been supported by French National Association of Technical Research (CIFRE 2020/0791).

\bibliographystyle{unsrt}

\end{document}